\documentclass[12pt]{article}
\usepackage{enumerate}
\usepackage[sort]{natbib}
\usepackage{graphicx}
\usepackage{xr}
\usepackage{dsfont}
\usepackage{amsmath}
\usepackage{mathtools}
\usepackage{xcolor}
\usepackage{hyperref}
\usepackage{cleveref}

\usepackage{stackengine}
\usepackage{verbatimbox}
\usepackage{setspace}      
\usepackage{multirow}
\usepackage{caption}
\usepackage{subcaption}
\usepackage{titlesec}
\usepackage{tabularx}
\newcommand{\blind}{1}
\usepackage{booktabs,array}
\usepackage{rotating}
\usepackage{amsthm}
\usepackage{amssymb}
\usepackage{arydshln}
\newtheorem{theorem}{Theorem}

\newtheorem{corollary}{Corollary}[theorem]

\usepackage[font=small]{caption}
\begin{document}
\def\spacingset#1{\renewcommand{\baselinestretch}%
{#1}\small\normalsize} \spacingset{1}
\if1\blind
{
  \title{\bf Smoothing Variances Across Time: Adaptive Stochastic Volatility}
  \author{Jason B. Cho\thanks{
    The authors gratefully acknowledge financial support from \textit{National Science Foundation grants OAC-1940124 and DMS-2114143}}\hspace{.2cm}\\
    Department of Statistics and Data Science, Cornell University\\
    and \\
    David S.\ Matteson \\
    Department of Statistics and Data Science, Cornell University}
  \maketitle
} \fi

\if0\blind
{
  \bigskip
  \bigskip
  \bigskip
  \begin{center}
    {\LARGE\bf Smoothing Variances Across Time: Adaptive Stochastic Volatility}
\end{center}
  \medskip
} \fi

\begin{abstract}
 We introduce a novel Bayesian framework for estimating time-varying volatility by extending the Random Walk Stochastic Volatility (RWSV) model with Dynamic Shrinkage Processes (DSP) in log-variances. Unlike the classical Stochastic Volatility (SV) or GARCH-type models with restrictive parametric stationarity assumptions, our proposed Adaptive Stochastic Volatility (ASV) model provides smooth yet dynamically adaptive estimates of evolving volatility and its uncertainty. We further enhance the model by incorporating a nugget effect, allowing it to flexibly capture small-scale variability while preserving smoothness elsewhere. We derive the theoretical properties of the global-local shrinkage prior DSP. Simulation studies demonstrate that ASV is highly robust to misspecification, consistently recovering the latent volatility structure across a wide range of data-generating processes. Furthermore, ASV's capacity to yield locally smooth and interpretable estimates facilitates a clearer understanding of the underlying patterns and trends in volatility. As an extension, we develop the Bayesian Trend Filter with ASV (BTF-ASV) which allows joint modeling of the mean and volatility with abrupt changes. Finally, our proposed models are applied to time series data from finance, econometrics, and environmental science, highlighting their flexibility and broad applicability.
\end{abstract}
\noindent%
\vfill
\newpage
\spacingset{1.75} 
\section{Introduction}\label{sec:intro}
Volatility of a time series quantifies deviations from the mean and is fundamental to understanding the underlying data-generating process. In finance, volatility estimation is essential for pricing, risk assessment, and asset management \citep{GARCH2,SVop}. In epidemiology, it aids in early outbreak detection and forecasting of new cases of diseases \citep{Epidem2,Epidem,epidem3}. Climate science measures volatility to study phenomena such as tornadoes, droughts, and rainfall \citep{climate1, climate2, climate3}. In engineering, volatility helps predict mechanical failures \citep{MH1, MH2}, while in hydrology, it informs understanding of streamflow variations \citep{geophysics4, geophysics1, geophysics2, geophysics3}.

\begin{figure}[!ht]
    \centering 
    \begin{subfigure}[b]{0.32\textwidth}
      \includegraphics[width=\linewidth]{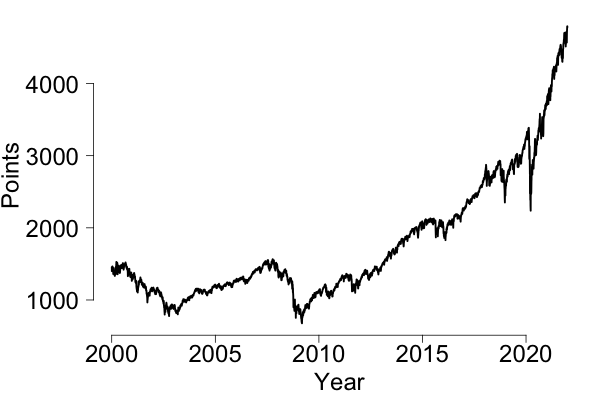}
      \caption{S\&P 500 Index}
      \label{fig:motive1}
    \end{subfigure}
    \begin{subfigure}[b]{0.32\textwidth}
      \includegraphics[width=\linewidth]{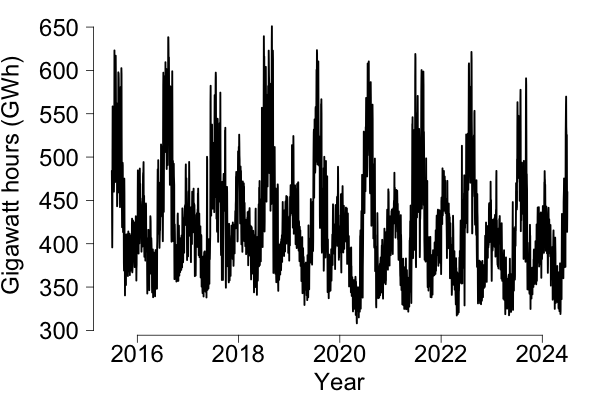}
      \caption{Electricity}
      \label{fig:motive2}
    \end{subfigure}
    \begin{subfigure}[b]{0.32\textwidth}
      \includegraphics[width=\linewidth]{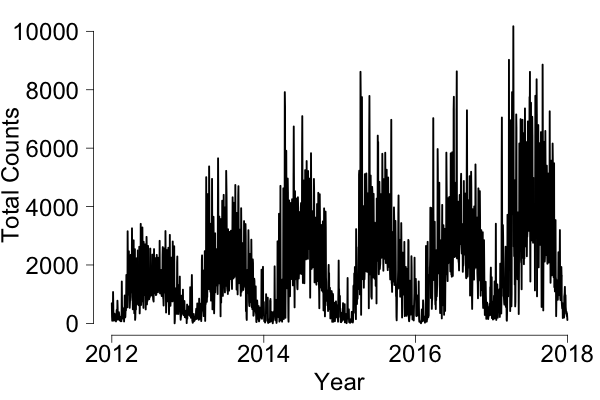}
      \caption{Bike Rentals}
      \label{fig:motive3}
    \end{subfigure}\hfil 
    \begin{subfigure}[b]{0.32\textwidth}
      \includegraphics[width=\linewidth]{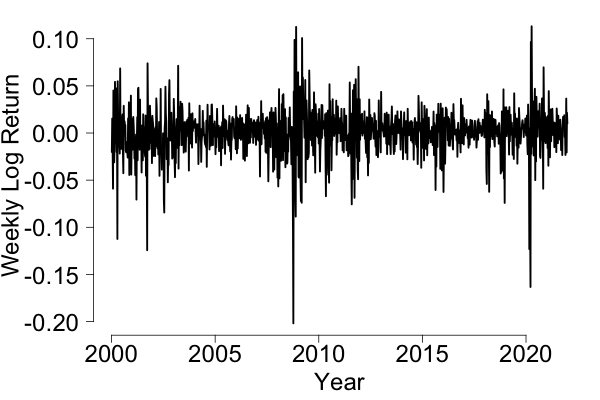}
      \caption{Weekly Log Returns}
      \label{fig:motive1b}
    \end{subfigure}
    \begin{subfigure}[b]{0.32\textwidth}
      \includegraphics[width=\linewidth]{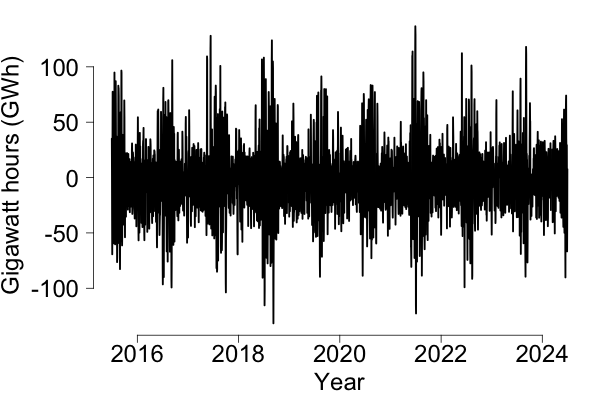}
      \caption{Electricity (mean-centered)}
      \label{fig:motive2b}
    \end{subfigure}
    \begin{subfigure}[b]{0.32\textwidth}
      \includegraphics[width=\linewidth]{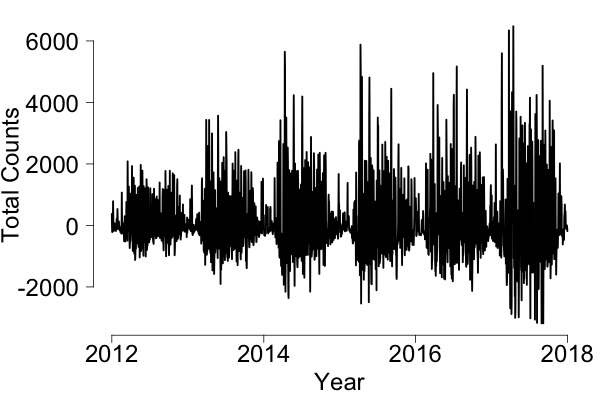}
      \caption{Bike Rentals (mean-centered)}
      \label{fig:motive3b}
    \end{subfigure}\hfil 
\caption{\Cref{fig:motive1,fig:motive2,fig:motive3} display the S\&P 500 index from 2001-01-01 to 2021-12-31, the daily total electricity demand in New York State from 2015-07-01 to 2024-06-30, and the daily bike rental counts via Capital Bikeshare by non-members in the Washington metropolitan area from 2012-01-01 to 2017-12-31, respectively. \Cref{fig:motive1b,fig:motive2b,fig:motive3b} show the corresponding mean-centered series: weekly log-returns for the S\&P 500 index with the average log return subtracted, and series for electricity demand and bike rentals mean-centered using a smoothing spline. The S\&P 500 index data are obtained using the \texttt{quantmod} package in \texttt{R} \citep{quantmod} from Yahoo Finance \citep{yahoo_finance_data}, the electricity demand data from the New York Independent System Operator (NYISO) \citep{nyis_electricity}, and the bike rental data from Capital Bikeshare website \citep{capital_bikeshare_data}.}
\label{fig:Motivation}
\end{figure}
Classical volatility models include AutoRegressive Conditional Heteroskedasticity (ARCH) \citep{ARCH}, Generalized ARCH (GARCH) \citep{GARCH2}, and Stochastic Volatility (SV) \citep{SVop,SVop2,SVop3}. They all make the stationarity assumption that the unconditional variance is constant. However, a growing body of econometric literature points to evolving volatility patterns in stock markets around the world \citep{NS2,NS3,UK,taiwan,chinese,SVRW_w}. Moreover, the stationarity assumption is often unrealistic in fields like climate science and epidemiology, due to climate change and seasonal or outbreak-driven dynamics \citep{climate1, epidem3}.

\Cref{fig:Motivation} illustrates evolving volatility patterns in three real-world datasets: weekly log-returns of the S\&P 500 index, daily electricity demand in New York State, and daily bike rental counts via Capital Bikeshare. All three exhibit nonstationary volatility patterns. In \Cref{fig:motive1,fig:motive1b}, persistent periods of high and low volatility reflect changing market conditions. In \Cref{fig:motive2,fig:motive2b,fig:motive3,fig:motive3b}, both the electricity and bike data show seasonal cycles in both mean and volatility, with elevated volatility during high-demand periods.

Existing approaches address such patterns by incorporating hidden Markov switching (HMS) structure into existing SV and GARCH models \citep{MSe1, MSe2, MS4, MSARCH, MSGarch, MSGARCH2, MS1, MSSV}. However, HMS models suffer from three major limitations: they (1) require the specification of the number of regimes, which is typically unknown; (2) are computationally intensive, as pointed out in \citet{MSefficiency1} and \citet{MSefficiency2}; and (3) impose a discrete regime-switching structure, which may not capture ever-evolving volatility dynamics. For example, \Cref{fig:motive3b} reveals a seasonal increase in volatility for bike rentals, with the magnitude of these fluctuations growing over time. Specifically, volatility in 2012 is substantially lower than in 2017. Such a pattern is difficult to reconcile with a model that is based on a fixed number of regimes, as it reflects a continuous evolution of volatility rather than discrete shifts.

We introduce the Adaptive Stochastic Volatility (ASV) model as an extension of the Random Walk Stochastic Volatility (RWSV) model \citep{SVRW1, SVRW2}. RWSV assumes that the log-variance of the observed process follows a Gaussian random walk with a constant variance. In contrast, ASV introduces two significant modifications to RWSV: it (1) incorporates a time-varying variance for the log-variance increments; and (2) adopts a global-local shrinkage prior for adaptability. For the shrinkage prior, we specifically highlight the Dynamic Shrinkage Processes (DSP) by \citet{dsp}, as it provides a flexible framework for adaptive shrinkage in time series. To further improve robustness, we consider a nugget variant of ASV, which adds an additional Gaussian noise term to capture local fluctuations not explained by the smooth latent process. 

The standout feature of our proposed model is its local adaptability, effectively estimating volatility in the presence of both gradual and abrupt changes in the volatility process, with notable smoothness in between. It exhibits robustness against model misspecification, enabling accurate recovery of the latent volatility structure across diverse data-generating processes. In addition, ASV's smooth volatility estimates provide clearer insights into the underlying patterns and trends. The model also naturally highlights periods of heightened volatility variation, offering additional interpretability in identifying structural shifts.

The remainder of the paper is organized as follows. Section~\ref{sec:model} introduces DSP, presents its theoretical properties, and describes the proposed ASV model. Section~\ref{sec:gibbs} outlines the data augmentation strategy for efficient Gibbs sampling, with full conditionals provided in Appendix B. Section~\ref{sec:simulationstudy} reports simulation results, and Section~\ref{sec:empiricalstudy} presents empirical analyses on data sets shown in Figure~\ref{fig:Motivation}. In section~\ref{sec:DSPmv}, the Bayesian Trend Filter with ASV (BTF-ASV), which jointly estimates the time-varying means and the variances, is proposed as an extension and applied to the sunspot data \citep{sunspot}.


\section{Methodology}\label{sec:model}
The goal of this section is to present the proposed ASV model, which captures evolving volatility with a global-local shrinkage prior. We begin with the RWSV model, the baseline formulation with a constant innovation variance. We then introduce an intermediate extension with a local-only shrinkage prior. We subsequently describe the DSP, a global-local shrinkage prior. Finally, we present the full ASV model and its nugget extension.

\subsection{The Random Walk Stochastic Volatility Model}\label{sec:rwsv}
We begin by reviewing the RWSV model \citep{SVRW1, SVRW2}. Consider a zero mean process with $T$ observations, $\{y_{t}\}_{t=1}^{T}$, and its log-variance term $\{h_{t}\}_{t=1}^{T}$. In the SV framework, the observations $y_t$ follow a normal distribution whose log-variance $h_t$ evolves according to a first-order autoregression. RWSV is a special case of SV in which the autoregressive coefficient is fixed at 1. The model is defined as:
\begin{equation}
    \begin{aligned}
    	&y_t = \exp\{h_{t}/2\}\epsilon_{t} & \quad & [\epsilon_{t}]\stackrel{iid}{\sim} N(0,1), \\ 
    	& [\Delta h_{t}|\sigma^2_h] \sim N(0, \sigma^2_{h}) &\quad & [\sigma^2_h] \sim \pi(\sigma^2_h).
    \end{aligned}\label{model:sv}
\end{equation}
In RWSV, $h_{t}$ is governed by a time-invariant variance term $\sigma_{h}^2$, which determines the degree of variation between successive log-variances. A large $\sigma_{h}^2$ implies high probability of significant changes in the $h_{t}$ process, whereas a small $\sigma_{h}^2$ indicates a high probability of minimal to no changes in $h_{t}$. In a frequentist framework, $\sigma_{h}$ is treated as a fixed parameter, often estimated via quasi-maximum likelihood \citep{SVRW1,SVRW2}. A more recent study by \citet{SVRW_w} employs a Bayesian framework for estimation with a non-informative inverse-gamma prior for $\pi(\sigma^2_h)$.

\subsection{Stochastic Volatility with Local Only Shrinkage}\label{sec:bayesianlasso}
RWSV may fit poorly when the time series exhibits both significant and minimal changes, as it assumes a time-invariant innovation variance $\sigma^2_h$. Ideally, we would prefer $\sigma^2_h$ to adapt over time, remaining small during stable periods and increasing during periods of abrupt changes in $h_t$. A natural approach to address this limitation is to introduce sparsity in the variance process. Under a frequentist framework, sparsity in $\Delta h_t$ may be attained by imposing a penalty, such as the $\ell_1$ penalty used in LASSO \citep{LASSO}. In a Bayesian setting, near-sparsity can be imposed via heavy-tailed priors on the variance parameter. Motivated by the Bayesian LASSO \citep{BayesianLasso} and the scale-mixture Gaussian representation of the Laplace distribution \citep{laplace_scale}, we propose an intermediate model, RWSV with Bayesian LASSO (RWSV-BL):
\begin{equation}
    \begin{aligned}
        &y_{t} = \exp\{h_t/2\}\epsilon_t & &[\epsilon_t] \stackrel{iid}{\sim}N(0,1),\\
        &[\Delta h_t|\sigma^2_{h,t}] \sim N(0, \sigma^2_{h,t}) & \quad &[\sigma^2_{h,t}|\Lambda] \stackrel{iid}{\sim} \mathrm{Exp}\{(2\Lambda^2)^{-1}\},
    \end{aligned}\label{model:rwsv-bl}
\end{equation}
where $\Lambda > 0$ is a hyperparameter controlling the prior concentration. Following \citet{BayesianLasso}, we place a Gamma prior on $\Lambda^2$ for full Bayesian inference. Under RWSV-BL, $\sigma^2_{h,t}$ induces shrinkage by encouraging most increments to be small while still allowing occasional large deviations.

\subsection{Dynamic Shrinkage Processes and their Properties}\label{sec:dsp_prop}
Near-sparsity may also be achieved through global-local shrinkage priors. These priors introduce a global parameter that controls the overall degree of shrinkage across time and local parameters that allow individual components to deviate when supported by the data. Given a sequence of parameters $\{\Theta_t\}_{t=1}^{T}$, they take the following hierarchical form:
\[
\Theta_t \sim N(0, \tau^2 \lambda_t^2), \qquad
\tau \sim \pi(\tau), \quad \lambda_t \sim \pi(\lambda_t),
\]
where $\tau$ governs global shrinkage and $\lambda_t$ captures local flexibility. This class of priors encourages most parameters to be strongly regularized while permitting occasional large deviations, making it well suited for volatility estimation, where periods of stability are occasionally interrupted by abrupt shifts. This structure naturally extends the SV formulation by placing the prior on the $k$th-order difference of the log-variance process, $\{\Delta^k h_{t}\}_{t=1}^{T}$, so that its variance decomposes as $\sigma^2_{h,t} = \tau^2 \lambda_t^2$.

Global-local shrinkage priors have been extensively studied in the context of high-dimensional Gaussian models, with notable examples including the horseshoe prior~\citep{horseshoe_theory}, the horseshoe+ prior~\citep{horseshoe_P}, and the triple-gamma prior~\citep{tgamma}. For time-series modeling, several related dynamic shrinkage formulations have been proposed. \citet{dsptype1} and \citet{dsptype2} extend horseshoe-type priors using heavy-tailed $t$ distributions and fusion-based structures. These priors share the same underlying motivation as the DSP, combining global regularization with locally adaptive shrinkage for recovering the underlying temporal structure. 

Among these priors, DSP offers a particularly flexible formulation that generalizes several existing priors and naturally incorporates temporal dependence. DSP has shown strong versatility across various applications. \citet{abco} integrate DSP into a Bayesian dynamic linear model to estimate change points and detect outliers. \citet{dspCount} apply it to develop the negative binomial Bayesian trend filter (NB-BTF), a method for smoothing integer-valued time series. Additionally, \citet{dspDrift} combine the Bayesian trend filter with DSP and a machine-learning-based regularization method to effectively distinguish micro-level drifts from macro-level shifts.

A defining feature of DSP is its autoregressive structure on its log-variance process. It enables locally adaptive shrinkage through heavy-tailed innovations around a persistent mean process. Let $v_t := \log(\sigma^2_{h,t}) = \log(\tau^2\lambda_t^2)$. $v_{t} \mid \mu,\phi \sim DSP(a, b, \mu, \phi)$ is defined by:
\begin{align}
	&v_{t} = \mu + \phi(v_{t-1}-\mu) + \eta_{t} & [\eta_{t}] \stackrel{iid}{\sim} Z(a,b,0,1),\label{eq:dsp}
\end{align}
where \( Z \)-distribution \citep{zdist} has density:
$$
f(z|a,b,\mu_{z},\sigma_{z}) = (\sigma_z \mathrm{B}(a,b))^{-1}\bigg(\exp\bigg\{\frac{z-\mu_z}{\sigma_z}\bigg\}\bigg)^{a}\bigg(1+\exp\bigg\{\frac{z-\mu_z}{\sigma_z}\bigg\}\bigg)^{-(a + b)}.
$$
For fully Bayesian inference, we place priors on $\mu$ and $\phi$, and describe them in detail in Section~\ref{sec:gibbs}. 

Under DSP, the global parameter is defined as $\tau = \exp\{\mu(1-\phi)/2\}$, and the local shrinkage parameter as $\lambda_t = \exp\{(\phi v_{t-1} + \eta_t)/2\}$. The coefficient $|\phi| <1$ controls temporal dependence in local shrinkage. When $\phi$ is close to 1, the process exhibits strong positive dependence: strong shrinkage tends to be followed by continued strong shrinkage, and weak shrinkage by weak shrinkage. When $\phi$ is close to -1, the process favors alternation between strong and weak shrinkage over time. 

The DSP includes several well-known priors as special cases. When $a = b = 1/2$ and the autoregressive coefficient $\phi = 0$, it reduces to the horseshoe (HS)  prior \citep{horseshoe_theory}, since $\lambda_{t} = \exp\{\eta_t/2\} \sim C^{+}(0,1)$. Keeping $a = b = 1/2$ but assigning $\phi$ its own prior (e.g., Beta) introduces temporal dependence in shrinkage, yielding the Dynamic Horseshoe (DHS) prior \citep{dsp}. Other notable cases include the Strawderman–Berger prior when $a = 1/2$ and $b = 1$ \citep{strawderman}, the normal–exponential–gamma prior when $a = 1$ and $b > 2$ \citep{griffin2005}, and the Jeffreys prior with $a = b \rightarrow 0$ \citep{figueiredo}.

While DSP has an appealing formulation and a clear connection to well-known priors, its theoretical properties remain incomplete. In particular, DSP distinguishes itself from other global-local shrinkage priors by introducing temporal dependence in the local shrinkage parameter \(\lambda_t\). Due to the time dependence, the original work by \citet{dsp} focuses on the conditional behavior, leaving the properties of the marginal distribution unexplored. We address this gap by deriving new results on the stationary distribution of DSP. 

Define $z_{\ell,t} := \phi^{\ell} \eta_{t-\ell}$. The infinite-order moving average representation of $v_{t}$ is given by $v_{t} = \mu + \phi(v_{t-1} - \mu) + \eta_{t} =  \mu + \sum_{\ell=0}^{\infty} z_{\ell,t}$, where $\eta_t \stackrel{iid}{\sim}Z(1/2,1/2,0,1)$. As shown in \citet{zdist}, $z_{\ell,t}$ follows a scaled hyperbolic secant distribution, with $\mathbb{E}(z_{\ell,t}|\phi) = 0$ and $\mathrm{Var}(z_{\ell,t}|\phi) = (|\phi|^{\ell} \pi)^2$. Let $z_t:= \sum_{\ell=0}^{\infty} z_{\ell,t}$ denote the aggregated process.
\begin{theorem}
	The series $z_t$ converges almost surely if and only if $|\phi|< 1.$
\end{theorem}
\begin{proof}
	This result directly follows from Kolmogorov's three-series theorem. The variance $\sum_{\ell=0}^{\infty} \mathrm{Var}(z_{\ell,t} \mid \phi)$ converges if and only if $|\phi| < 1$. The exact derivation is provided in Appendix A.1.
\end{proof}
\begin{corollary}
If $|\phi| < 1$, then
\begin{align*}
	&\mathbb{E}
(z_{t}|\phi) = \mathbb{E}
\bigg(\sum_{\ell=0}^{\infty} z_{\ell,t}|\phi\bigg) = \sum_{\ell=0}^{\infty} \mathbb{E}
(z_{\ell,t}|\phi)= 0. \\
	&\mathrm{Var}(z_{t}|\phi) = \mathrm{Var}\bigg(\sum_{\ell=0}^{\infty} z_{\ell,t}|\phi\bigg) = \sum_{\ell=0}^{\infty}  \mathrm{Var}(z_{\ell,t}|\phi) =  \sum_{\ell=0}^{\infty}  \pi^2 \phi^{2\ell} =\frac{\pi^2}{1-\phi^{2}}.
\end{align*}
The characteristic function of $z_{\ell,t}$ is $\mathrm{sech}(\pi \phi^\ell t)$. Thus, the characteristic function, $g(t)$ for $\sum_{\ell=0}^{\infty} z_{\ell,t}$ would be the infinite product of the characteristic function of $z_{\ell,t}$.
\begin{align*}
	&g(t) = \prod_{\ell=0}^{\infty} \mathrm{sech}(\pi\phi^{\ell}t),  & -\frac{1}{2} < t < \frac{1}{2}.
\end{align*}
Similarly, for the moment generating function $M(t)$ is given by
\begin{align*}
	&M(t) = \prod_{\ell=0}^{\infty} \mathrm{sec}(\pi\phi^{\ell}t),  & -\frac{1}{2}< t < \frac{1}{2}.
\end{align*} 
\end{corollary}

\begin{theorem}
	If $|\phi| = 0.5$, $\eta_{t-\ell} \stackrel{iid}{\sim} Z(1/2,1/2,0,1)$, and $z_{\ell,t} := \phi^\ell \eta_{t-\ell}$, then $z_t := \sum_{\ell=0}^{\infty} z_{\ell,t} \stackrel{a.s}{\longrightarrow} \mathrm{Logistic}(0,2)$. 
\end{theorem}
\begin{proof}
	Derivations are explored in Appendix A.2.
\end{proof}
Thus, assuming $\mu = 0$ and $\phi = 0.5$ for simplicity, we have the following density functions for $v_{t}$, $\lambda_{t}:= \exp\{v_{t}/2\}$, and $\kappa_{t} := 1/(1 + \exp\{v_{t}\})$, corresponding to the stationary distribution of DSP: 
\begin{align*}
	&f(v_{t}) = \frac{1}{8} \mathrm{sech}^2\bigg(\frac{v_{t}}{4}\bigg), \\
	&f(\lambda_t) = \frac{1}{4\lambda_t} \mathrm{sech}^2\bigg(\frac{\log(\lambda_{t})}{2}\bigg) = \frac{1}{(1+\lambda_{t})^2}, \\
	&f(\kappa_{t}) = \frac{1}{8(\kappa_{t})(1-\kappa_{t})} \mathrm{sech}^2\bigg(\frac{1}{4}\log\bigg(\frac{1-\kappa_{t}}{\kappa_{t}}\bigg) \bigg)  = \frac{1}{2} \kappa_{t}^{-3/2}(1-\kappa_{t})^{-1/2}\bigg(1+ \sqrt{\frac{1-\kappa_{t}}{\kappa_{t}}}\bigg)^{-2}.
\end{align*}
\begin{figure}[ht]
    \centering 
    \begin{subfigure}[b]{0.4\textwidth}
      \includegraphics[width=\linewidth]{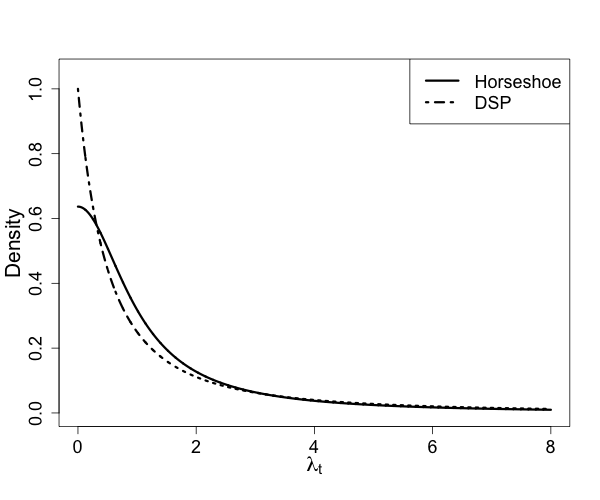}
      \caption{Prior densities on $\lambda_t$}
    \end{subfigure}
        \begin{subfigure}[b]{0.4\textwidth}
      \includegraphics[width=\linewidth]{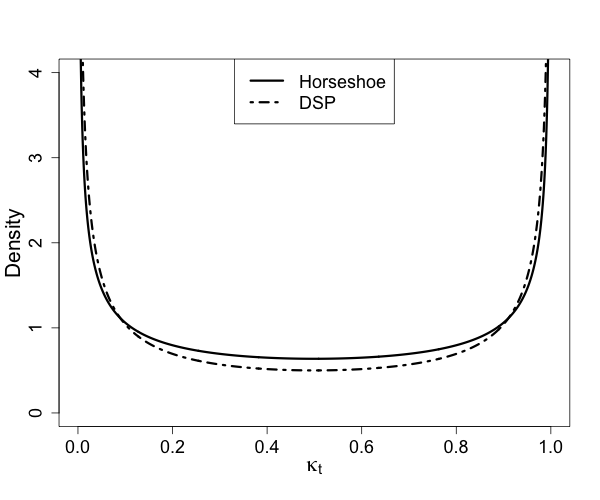}
      \caption{Prior densities on $\kappa_t$}
    \end{subfigure}
\caption{Prior densities of $\lambda_t$ and $\kappa_t$ under the horseshoe prior and the stationary distribution of Dynamic Shrinkage Process (DSP) with $\eta_{t} \stackrel{iid}{\sim} Z(1/2,1/2,0,1)$, $\phi = 1/2$ and $\mu = 0$.}
\label{fig:DSPvsHS}
\end{figure}
Figure~\ref{fig:DSPvsHS} compares the prior distribution of $\lambda_{t}$ and $\kappa_{t}$ under the horseshoe and DSP priors. The horseshoe prior assumes $\lambda_{t} \sim C^{+}(0,1)$ and $\kappa_{t} \sim \text{Beta}(1/2,1/2)$. The corresponding density functions, $g(\lambda_{t})$ and $g(\kappa_{t})$, are given by :
\begin{align*}
	&g(\lambda_{t})= \frac{2}{\pi(1+\lambda_t^2)}, & g(\kappa_{t})= \frac{1}{\pi \sqrt{\kappa_{t}(1-\kappa_{t})}}.
\end{align*}
For $\lambda_{t}$, the DSP density has heavier tails and a sharper spike at zero. Specifically, $f(\lambda_{t}) > g(\lambda_{t})$ when $\lambda_{t} \in \bigg(0,\frac{2 - \sqrt{(4-\pi)\pi}}{(\pi-2)}\bigg) \cup \bigg(\frac{2 + \sqrt{(4-\pi)\pi}}{(\pi-2)}, \infty\bigg)$. For $\kappa_{t}$, the DSP prior places more weight near 0 and 1 than the horseshoe. This behavior arises from the additional U-shaped term in $\kappa_t$, $(2\kappa_{t})^{-1} \bigg(1+ \sqrt{\frac{1-\kappa_{t}}{\kappa_{t}}} \bigg)^{-2}$, which pushes the mass of the distribution toward the boundaries. 

Finally, the marginal prior on $\{\Delta h_t\}_{t=1}^{T}$ is considered: 
\begin{theorem}
    Let $|\phi| = 0.5$, $\eta_{t-\ell} \stackrel{iid}{\sim} Z(1/2,1/2,0,1)$, $z_{\ell,t} := \phi^{\ell} \eta_{t-\ell}$, $v_t := \sum_{\ell=0}^{\infty} z_{\ell,t}$, $\lambda_t := \exp\{v_t/2\}$, and $\Delta h_t \sim N(0, \lambda_t^2)$. Then
	\begin{align*}
		&\lim_{\Delta h_t \to 0} f(\Delta h_t) = \infty, & \\
		&K_L \log\bigg(1+\frac{4}{(\Delta h_t)^2}\bigg) < f(\Delta h_t) < K_U \log\bigg(1+\frac{2}{(\Delta h_t)^2}\bigg), & |\Delta h_{t}| > 0
	\end{align*}
	where $K_U = 1/(2\sqrt{2\pi})$ and $K_L = 1/(8\sqrt{2\pi})$
\end{theorem}
\begin{proof}
	The proof is similar to Theorem 1 in \citet{horseshoe_theory}. We use the fact that $\forall x > 0$,
	$$ \frac{1}{2(1+x^{2})} \leq \frac{1}{(1+x)^2} \leq \frac{1}{(1+x^2)}.$$
	Derivation is detailed in Appendix A.3.
\end{proof}
The bounds for the marginal distribution of DSP are similar to the ones from the horseshoe with the only difference being the constant factor $K_L$ and $K_U$. Under the horseshoe prior, $K_U = 1/\sqrt{2\pi^3}$ and $K_L = 1/(2\sqrt{2\pi^3})$. Like horseshoe prior, DSP is also unbounded near the origin, which leads to super-efficiency in a sparse setting as shown in \citet{horseshoe_theory}.

\subsection{Adaptive Stochastic Volatility via Global-Local Shrinkage}
We propose ASV, which uses DSP explored in Section~\ref{sec:dsp_prop}, to estimate time-varying volatility. In ASV, the prior models the log-variance of the $k$th-order differences of $h_t$, where $\Delta^k h_t \mid \sigma^2_{h,t} \sim N(0, \sigma^2_{h,t})$. As previously defined, \( v_t := \log(\sigma^2_{h,t}) \) follows the autoregressive structure described in Equation~\ref{eq:dsp}. Unlike RWSV-BL, ASV incorporates both the global parameter, $\mu(1-\phi)$, and the local parameter, $\phi v_{t-1} + \eta_{t}$. The resulting hierarchical representation of ASV is given by:
\begin{equation}
    \begin{aligned}
    & y_{t} = \exp\{h_t/2\}\epsilon_t & &[\epsilon_t] \stackrel{iid}{\sim}N(0,1),\\
    & [\Delta^k h_{t}|\sigma^2_{h,t}] \sim N(0,\sigma_{h,t}^2) & & [\log(\sigma^2_{h,t})|\mu,\phi] \sim DSP(a,b,\mu,\phi).\\
\end{aligned} \label{eq:asv}   
\end{equation}

The differencing order $k$ plays a central role in controlling the model's smoothness. ASV can be viewed as a variance-process analogue of smoothing methods for the mean function, such as the Bayesian Trend Filter by \citet{btf} and \citet{dsp}, the Hodrick-Prescott filter by \citet{hp}, the $\ell_1$ trend filter by \citet{l1tf1} and \citet{l1tf2}, and the smoothing spline by \citet{Sspline}. Penalizing the $k$th-order difference encourages the underlying signal to be locally well-approximated by a polynomial of degree $k-1$: $k=1$ promotes piecewise constant behavior, $k=2$ encourages piecewise linearity, and higher-order penalties induce smoother trends. In our study, we use $k = 1$, consistent with traditional volatility models like GARCH and SV, which typically assume stationary dynamics with abrupt clustering. Nevertheless, higher-order differencing may be better suited for applications with more gradually evolving volatility.

In practice, volatility processes may exhibit local irregularities that are neither smooth nor abrupt. To account for this, we extend ASV by introducing a nugget effect, an additional noise term that absorbs small-scale local fluctuations. We refer to this specification as ASV with nugget:
\begin{equation}
    \begin{aligned}
    & y_{t} = \exp\{h_t/2\}\epsilon_t & &[\epsilon_t] \stackrel{iid}{\sim}N(0,1),\\
    & h_{t} = h^*_t + u_t & & [u_t|\sigma^2_{c}] \sim N(0,\sigma^2_{c}), \\
    & [\Delta^kh_{t}^*|\sigma^2_{h,t}] \sim N(0,\sigma^2_{h,t}) & & [\log(\sigma^2_{h,t})|\mu,\phi] \sim DSP(a,b,\mu,\phi).
\end{aligned} \label{eq:asv_nug}   
\end{equation}
It shares the same observation model as its non-nugget counterpart in Equation~\ref{eq:asv}, but differs in that the log-variance $h_t$ is decomposed into a smooth component $h_{t}^*$, and a local noise term, $u_t$, allowing the model to capture both structured and unstructured variation.

\section{Bayesian Inference via Data Augmentation}\label{sec:gibbs}
We briefly outline the Bayesian inference procedure for the ASV model described in Equation~\ref{eq:asv}. The observed data are denoted by $\boldsymbol{y} := (y_{1},\ldots,y_{T})'$, the latent log-variance process by $\boldsymbol{h} := (h_{1},\ldots, h_{T})'$ and its associated log-variance innovation process by $\boldsymbol{v} := (\log(\sigma^2_{h,1}),\ldots,\log(\sigma^2_{h,T}))'$. For the autoregressive parameter $\phi$, we assume a shifted Beta prior, $(\phi + 1)/2 \sim \mathrm{Beta}(10,2)$ \citep{dsp}. This prior favors positive $\phi$, reflecting the persistent nature of volatility dynamic, while discouraging the unrealistic oscillatory behavior that occurs when $\phi < 0.$ For the global parameter, we assume $\mu \sim Z(1/2, 1/2, 0, 1)$, which implies $\tau = \exp\{\mu(1 - \phi)/2\} \sim C^+(0,1)$ when $\phi = 0$. This recovers the same prior on the local parameter in \citet{horseshoe_theory}. Inference proceeds by sampling from the joint posterior distribution of $(\boldsymbol{h},\boldsymbol{v},\mu,\phi)$ conditional on $\boldsymbol{y}$ via a Gibbs sampling scheme that iteratively samples from the full conditional distributions. 

Sampling from the conditional posterior distribution of $\boldsymbol{h}$ in the SV model is challenging due to its nonlinear likelihood. While methods such as sequential Monte Carlo by \citet{sequentialMH} can sample \( h_t \) from the exact conditional posterior sequentially, this approach is computationally demanding. Instead, we adopt a quasi-likelihood approach based on the transformed observations $\boldsymbol{y^*} := (\log(y^2_{1}),\ldots, \log(y^2_{T}))'$, recasting the model into a linear system with a non-Gaussian error term: $y_{t}^* = h_{t} + \log(\epsilon_{t}^2)$, where $\epsilon_t \stackrel{iid}{\sim} N(0,1)$. 

The asymmetric error term $\log(\epsilon_{t}^2)$ is approccimated by a finite Gaussian mixture, originally proposed by \citet{kimetal} and later refined by \citet{omorietal} using a 10-component Gaussian mixture:
\begin{align*}
    &\log(\epsilon_{t}^2) \approx m_{j_{t}} + o_t & [o_t|j_t] \stackrel{iid}{\sim} N(0,w_{j_{t}} ^2),
\end{align*}
where $j_{t} \stackrel{iid}{\sim} \mathrm{Categorical}(\pi^{\mathrm{Omori}})$, with fixed weights $\pi^{\mathrm{Omori}} = (\pi_{1},\ldots,\pi_{10})$. Each latent index $j_t \in \{1,\ldots,10\}$ selects a component-specific mean $m_{j_t}$ and standard deviation $w_{j_t}$ from a pre-specified table \citep{omorietal}. This approximation transforms the quasi-likelihood into a conditionally Gaussian model, making posterior sampling significantly more efficient. Conditional on $j_t$, the full conditional distribution of $\boldsymbol{h}$ becomes a Gaussian likelihood with a structured Gaussian prior:
\begin{equation*}
    \begin{aligned}
        &y_t^* \approx h_{t} + m_{j_{t}} + o_{t} & &[o_{t}|j_{t}]\stackrel{ind}{\sim}N(0,w_{j_{t}}^2),\\
        &[\Delta^kh_{t}|v_t] \sim N(0,\exp(v_t)) & & [v_t|\mu,\phi] \sim DSP(a,b,\mu,\phi).
    \end{aligned}
\end{equation*}
The same Gaussian mixture approximation and data augmentation are applied to the latent innovation term $\omega_t := \Delta^k h_{t}$. Specifically, since $\Delta^kh_{t}|v_t \sim N(0,\exp(v_t))$, we can express 
	\begin{align*}
			&\omega_t = \exp\{v_t/2\}a_t & [a_t] \stackrel{iid}{\sim} N(0,1).
	\end{align*}
Therefore, $\log(\omega_t^2) := v_t + \log(a_t^2)$, where $\log(a_t^2)$ follows the same asymmetric distribution as $\log(\epsilon_t^2)$ for the process $\{h_t\}_{t=1}^{T}$. Consequently, the same 10-component Gaussian mixture approximation can be used, introducing a corresponding latent indicator variable $s_t$.

The evolution equation of $v_{t}$ in Equation~\ref{eq:dsp} may be interpreted as the SV model with a heavy-tailed error distribution \( \eta_{t} \sim Z(a, b, 0, 1) \). We leverage the P\'{o}lya-Gamma scale mixture representation of the \( Z \)-distribution \citep{zdist}, which states:
\begin{align*}
    &[\eta_t|\xi_t] \sim N(\xi_t^{-1}(a - b)/2,\xi_{t}^{-1}) & [\xi_t] \sim PG(a+b,0).
\end{align*}
Combined with the approximation on $\log(\omega_t^2) = \log((\Delta^k h_{t})^2)$, the log-variance evolution becomes conditionally Gaussian:
\begin{equation*}
    \begin{aligned}
    &\log(\omega_t^2)\approx v_{t} + m_{s_{t}} + r_{t} & & [r_{t}|s_{t}] \stackrel{ind}{\sim} N(0,w_{s_{t}}^2),\\ 
      &v_{t} = \mu + \phi(v_{t-1} - \mu) + \eta_{t} & &[\eta_t|\xi_t] \stackrel{ind}{\sim} N(\xi_t^{-1}(a - b)/2,\xi_{t}^{-1}).   
    \end{aligned}
\end{equation*}
 Similarly, $\xi_{\mu}\sim PG(1,0)$ is introduced to represent $\mu \sim Z(1/2,1/2,0,1)$ as $\mu\mid\xi_{\mu} \sim N(0,\xi_{\mu}^{-1})$. Sampling of $\boldsymbol{v}$ follows the AWOL sampler by \citet{Kastner_2014}, and P\'{o}lya-Gamma variables are drawn using a truncated an infinite sum as in \citep{polyagamma}.

In addition to the four existing parameters $(\boldsymbol{h},\boldsymbol{v}, \mu, \phi)$,
four additional parameters are introduced: the mixture indicators $\boldsymbol{j}:= (j_{1},\ldots, j_{T})'$ and $ \boldsymbol{s} := (s_{1},\ldots, s_{T})'$, and the P\'{o}lya-Gamma variables: $\boldsymbol{\xi}:= (\xi_{1} \ldots \xi_{T})'$ and $\xi_{\mu}$. Detailed derivations of the full conditional distributions are discussed in the Appendix B. For the nugget model described in Equation~\ref{eq:asv_nug}, the same approximation and data augmentation schemes are applied. An additional Gaussian layer, $h_t \mid h_t^*,\sigma^2_c \sim N(h_t^*,\sigma^2_c)$, is introduced, and the prior is placed on $\Delta^k h^*_{t}$ instead of $\Delta^k h_t$. The model remains conditionally Gaussian, and the structure of the Gibbs sampler is largely unchanged.
\section{Simulation Study}\label{sec:simulationstudy}
\subsection{Set-up}
The simulation study evaluates the performance of the proposed ASV model against several existing methods, including the SV, Markov-Switching SV with two regimes (MSSV2), RWSV, GARCH, and Markov-Switching GARCH with two regimes (MSGARCH2). A local-only shrinkage model, RWSV-BL, is also included for comparison. Four variants of ASV are considered: ASV with the horseshoe prior (ASV-HS), ASV with the dynamic horseshoe prior (ASV-DHS), and their respective versions with a nugget effect (ASV-HS-N and ASV-DHS-N). ASV-HS corresponds to the special case of Equation~\ref{eq:asv} with $a = b = 1/2$ and $\phi = 0$, while ASV-DHS generalizes this by allowing $\phi$ to be estimated from the data. All ASV models use $k = 1$, applying shrinkage to the first-order differences of the latent log-variance process. This choice reflects the empirical observation that volatility in SV- and GARCH-type models typically exhibits clustering rather than gradual changes in slope.

Eight DGPs are considered: DGPs 1-3 are based on SV models with one to three regimes, while DGPs 4–6 are based on GARCH models with one to three regimes. To ensure that the simulated data reflect realistic volatility dynamics, the parameters for each DGP are calibrated using estimates obtained from empirical data. Specifically, DGPs 1 and 4 use parameters estimated from the S\&P 500 data (\Cref{fig:motive1}), DGPs 2 and 5 from the electricity data (\Cref{fig:motive2}), and DGPs 3 and 6 from the bike rental data (\Cref{fig:motive3}). DGP 7 features sinusoidal log-variance with random coefficients, and DGP 8 features a piecewise constant log-variance with three to five random breakpoints. For each DGP, 1,000 sample paths are simulated. Detailed parameterization of the DGPs, including the transition probabilities, are provided in Appendix C.

Model performance is evaluated using three metrics: Mean Absolute Error (MAE), Empirical Coverage (EC), and Mean Credible Interval Width (MCIW):
{\small
\begin{align*}
    MAE &= \frac{1}{T}\sum_{t=1}^{T}|\sigma_t - \hat{\sigma_t}|, &  
    EC &= \frac{1}{T}\sum_{t=1}^{T} \mathds{1}_{(\hat{\sigma}_{t,0.05},\hat{\sigma}_{t,0.95})}(\sigma_t), &
    MCIW &= \frac{1}{T}\sum_{t=1}^{T} (\hat{\sigma}_{t,0.95} - \hat{\sigma}_{t,0.05}).
\end{align*}}
The MAE measures the accuracy of the volatility estimate by calculating the average absolute difference between the true volatility, \(\sigma_t\), and the model's estimate, \(\hat{\sigma_t}\). For frequentist methods, such as GARCH and MSGARCH2, the point estimate corresponds to the maximum likelihood estimate, whereas for Bayesian methods, it corresponds to the posterior mean. The EC assesses the reliability of each model's uncertainty quantification by calculating the proportion of true volatility values that fall within the model’s 90\% credible interval. Finally, the MCIW evaluates the precision of this uncertainty quantification, with narrower credible intervals indicating higher precision provided that nominal coverage is achieved. Because GARCH and MSGARCH2 provide deterministic estimates of conditional volatility given the fitted parameters and data, MCIW and EC are not reported for these models.

All models are implemented in the {\tt R} language \citep{Rlang}. Specifically, the SV model is implemented using the {\tt stochvol} package \citep{Stochvol}; MSSV2, as specified in \citet{MSSVmine}, and RWSV are implemented by the authors, as no readily available {\tt R} packages exist. The GARCH(1,1) model is fitted using the {\tt fGarch} package \citep{fGarch}, and MSGARCH2 is fitted using the {\tt MSGARCH} package \citep{MSGARCHP}. To fit RWSV-BL, the {\tt dsp} package by \citet{dsp} and the {\tt genlasso} package by \citet{genlasso} are used. The {\tt dsp} package is extended by the authors to implement ASV. For Bayesian models, 5,000 posterior samples are retained after 20,000 burn-in iterations.
\subsection{Results}
\begin{sidewaystable}[!htbp]
\centering
\scriptsize
\setlength{\tabcolsep}{1.5pt} 
\begin{tabular}{ll*{8}{c}}
\toprule
 & & \multicolumn{8}{c}{\textbf{DGP ID}} \\
\cmidrule(l){3-10}
 & & 1 & 2 & 3 & 4 & 5 & 6 & 7 & 8 \\
 & \textbf{True Model}
   & SV (1 regime) & SV (2 regimes) & SV (3 regimes)
   & GARCH (1 regime) & GARCH (2 regimes) & GARCH (3 regimes)
   & Sinusoidal & Piecewise \\
\midrule

\multirow{10}{*}{\textbf{MAE}}
& SV & \textbf{4.267e-03 (4.581e-04)} & 16.6293 (3.9431) & 2.1691 (0.4344) & 0.0245 (0.0055) & 4.1234 (0.4709) & 1.1816 (0.261) & 1.1831 (0.9398) & 2.0796 (0.1713) \\
& MSSV2 & \textbf{4.302e-03 (4.781e-04)} & \textbf{6.7195 (1.9145)} & 2.0253 (0.3813) & 0.0248 (0.0056) & \textbf{3.0737 (0.7494)} & 1.0478 (0.2646) & 1.0434 (0.8259) & 1.4738 (0.2381) \\
& GARCH & 5.982e-03 (7.158e-04) & 34.3626 (8.3057) & 3.5958 (0.7481) & \textbf{0.0047 (0.0026)} & 5.6249 (0.6527) & 1.7023 (0.3995) & 2.7572 (2.2338) & 4.1509 (0.3525) \\
& MSGARCH2 & 5.909e-03 (7.031e-04) & 17.8737 (1.866) & 2.8629 (0.5375) & \textbf{0.0063 (0.0036)} & 3.8062 (0.6359) & 1.2492 (0.3289) & 2.1427 (1.6771) & 2.7224 (0.7748) \\
& RWSV & 4.359e-03 (4.699e-04) & 16.3886 (3.8574) & 2.1386 (0.421) & 0.0251 (0.0054) & 4.0889 (0.4599) & 1.1638 (0.254) & 1.1668 (0.925) & 2.0301 (0.1692) \\
& RWSV-BL & 5.223e-03 (5.582e-04) & 19.4377 (4.6705) & 2.5723 (0.5261) & 0.0319 (0.0068) & 5.742 (0.6862) & 1.6914 (0.4265) & 1.42 (1.0283) & 1.9933 (0.1658) \\
\noalign{\vskip 4pt}
\cdashline{2-10}[4pt/2pt]
\noalign{\vskip 4pt}
& ASV-HS & 4.562e-03 (4.918e-04) & 13.0979 (2.836) & 1.4824 (0.3262) & 0.026 (0.0056) & 3.4967 (0.5326) & \textbf{0.9538 (0.23)} & \textbf{0.9348 (0.7365)} & 1.2409 (0.1392) \\
& ASV-DHS & 4.601e-03 (4.976e-04) & \textbf{12.4654 (2.6887)} & \textbf{1.4365 (0.3183)} & 0.0262 (0.0056) & \textbf{3.4395 (0.542)} & \textbf{0.9551 (0.2301)} & 0.9426 (0.7442) & \textbf{1.2209 (0.1385)} \\
& ASV-HS-N & 4.562e-03 (4.940e-04) & 13.1351 (2.8391) & 1.4866 (0.3266) & 0.0262 (0.0056) & 3.5331 (0.5424) & 0.9603 (0.2311) & \textbf{0.9358 (0.7373)} & 1.2449 (0.1386) \\
& ASV-DHS-N & 4.592e-03 (4.977e-04) & 12.5345 (2.6921) & \textbf{1.4475 (0.3189)} & 0.0263 (0.0056) & 3.4831 (0.545) & 0.9619 (0.2316) & 0.9427 (0.7423) & \textbf{1.2242 (0.1382)} \\
\midrule

\multirow{10}{*}{\textbf{MCIW}}
& SV & 0.01787 (0.002083) & 93.5561 (22.6665) & 10.4085 (2.3493) & 0.1055 (0.0207) & 18.7136 (2.4446) & 6.1095 (1.4139) & 6.9224 (5.4048) & 10.5937 (0.8248) \\
& MSSV2 & 0.01745 (0.002244) & \textbf{35.6871 (10.954)} & 9.3695 (2.015) & 0.1003 (0.0202) & 20.0305 (3.6871) & \textbf{4.4687 (1.2165)} & 5.3407 (4.2339) & 6.6075 (1.0614) \\
& GARCH & - & - & - & - & - & - & - & - \\
& MSGARCH2 & - & - & - & - & - & - & - & - \\
& RWSV & 0.01527 (0.001879) & 91.3216 (22.0039) & 9.8339 (2.211) & 0.0955 (0.0215) & 16.5878 (2.0431) & 5.8831 (1.3353) & 6.801 (5.3228) & 10.0901 (0.7914) \\
& RWSV-BL & 0.02885 (0.002798) & 114.9675 (27.5597) & 14.4477 (2.9798) & 0.1808 (0.0331) & 33.4065 (3.6979) & 10.0222 (2.5165) & 8.4838 (6.0041) & 10.6267 (0.7472) \\
\noalign{\vskip 4pt}
\cdashline{2-10}[4pt/2pt]
\noalign{\vskip 4pt}
& ASV-HS & \textbf{0.01415 (0.001847)} & 65.2994 (13.9958) & \textbf{6.6685 (1.3651)} & \textbf{0.0903 (0.0217)} & \textbf{14.2277 (1.8101)} & 4.4709 (0.9961) & \textbf{4.2954 (3.2816)} & \textbf{6.4435 (0.4201)} \\
& ASV-DHS & \textbf{0.01396 (0.001849)} & \textbf{60.8098 (12.7929)} & \textbf{6.4654 (1.2943)} & \textbf{0.0892 (0.0216)} & \textbf{13.9008 (1.763)} & \textbf{4.4124 (0.9761)} & \textbf{4.3191 (3.3025)} & \textbf{6.3782 (0.4163)} \\
& ASV-HS-N & 0.01736 (0.002667) & 68.8245 (14.8542) & 7.3104 (1.4842) & 0.0948 (0.0221) & 16.2665 (2.2526) & 4.8062 (1.0979) & 4.6034 (3.4542) & 6.6306 (0.4301) \\
& ASV-DHS-N & 0.01742 (0.002744) & 65.1302 (13.6983) & 7.1792 (1.4312) & 0.0943 (0.022) & 16.0765 (2.1986) & 4.7761 (1.0848) & 4.6343 (3.4943) & 6.5572 (0.4303) \\
\midrule

\multirow{10}{*}{\textbf{EC}}
& SV & \textbf{0.9037 (0.0276)} & 0.9584 (0.0125) & 0.9134 (0.0162) & 0.9323 (0.0136) & 0.9056 (0.0251) & 0.9322 (0.0152) & 0.9764 (0.0114) & \textbf{0.9091 (0.0107)} \\
& MSSV2 & \textbf{0.89 (0.04476)} & 0.9482 (0.0352) & \textbf{0.9079 (0.0234)} & \textbf{0.9061 (0.036)} & 0.9618 (0.0522) & \textbf{0.9059 (0.0457)} & 0.9543 (0.0251) & 0.9539 (0.0237) \\
& GARCH & - & - & - & - & - & - & - & - \\
& MSGARCH2 & - & - & - & - & - & - & - & - \\
& RWSV & 0.8326 (0.04226) & 0.9555 (0.0132) & \textbf{0.9002 (0.018)} & \textbf{0.8884 (0.021)} & 0.8665 (0.0314) & \textbf{0.9261 (0.0168)} & 0.9758 (0.0118) & \textbf{0.8973 (0.0112)} \\
& RWSV-BL & 0.9637 (0.01033) & 0.9696 (0.0094) & 0.9593 (0.0108) & 0.9675 (0.0101) & 0.9685 (0.0096) & 0.9686 (0.0097) & 0.975 (0.0093) & 0.9422 (0.0107) \\
\noalign{\vskip 4pt}
\cdashline{2-10}[4pt/2pt]
\noalign{\vskip 4pt}
& ASV-HS & 0.7652 (0.04822) & \textbf{0.9398 (0.0203)} & 0.9326 (0.0279) & 0.8512 (0.0349) & \textbf{0.9024 (0.0395)} & 0.9306 (0.0274) & \textbf{0.9275 (0.0295)} & 0.9569 (0.0168) \\
& ASV-DHS & 0.7526 (0.04992) & \textbf{0.9371 (0.0228)} & 0.9332 (0.0291) & 0.8398 (0.0367) & \textbf{0.9027 (0.0412)} & 0.9264 (0.029) & \textbf{0.9264 (0.0296)} & 0.9583 (0.017) \\
& ASV-HS-N & 0.8587 (0.04874) & 0.9489 (0.0178) & 0.9502 (0.0231) & 0.8711 (0.0291) & 0.9269 (0.0315) & 0.9452 (0.0229) & 0.9498 (0.0245) & 0.9614 (0.0156) \\
& ASV-DHS-N & 0.8567 (0.05088) & 0.9493 (0.0193) & 0.9531 (0.023) & 0.8643 (0.0303) & 0.9288 (0.032) & 0.9431 (0.0238) & 0.9494 (0.0245) & 0.963 (0.0155) \\
\bottomrule
\end{tabular}
\caption{Model performance under each data-generating process (DGPs 1–8), evaluated using: MAE (mean absolute error), MCIW (mean 90\% credible interval width), and EC (empirical 90\% coverage). Standard deviations are shown in parentheses. For each DGP, the two models with the lowest average MAE, the model closest to the nominal 90\% coverage, and the model with the narrowest average MCIW are highlighted in bold. MCIW and EC are not reported for GARCH and MSGARCH2, as these frequentist models yield deterministic estimates of conditional volatility given the fitted parameters and data.}
\label{tab:wide_results}
\end{sidewaystable}
\begin{sidewaysfigure}[!htbp]
    \centering
    \begin{subfigure}[b]{0.24\textwidth}
      \includegraphics[width=\linewidth]{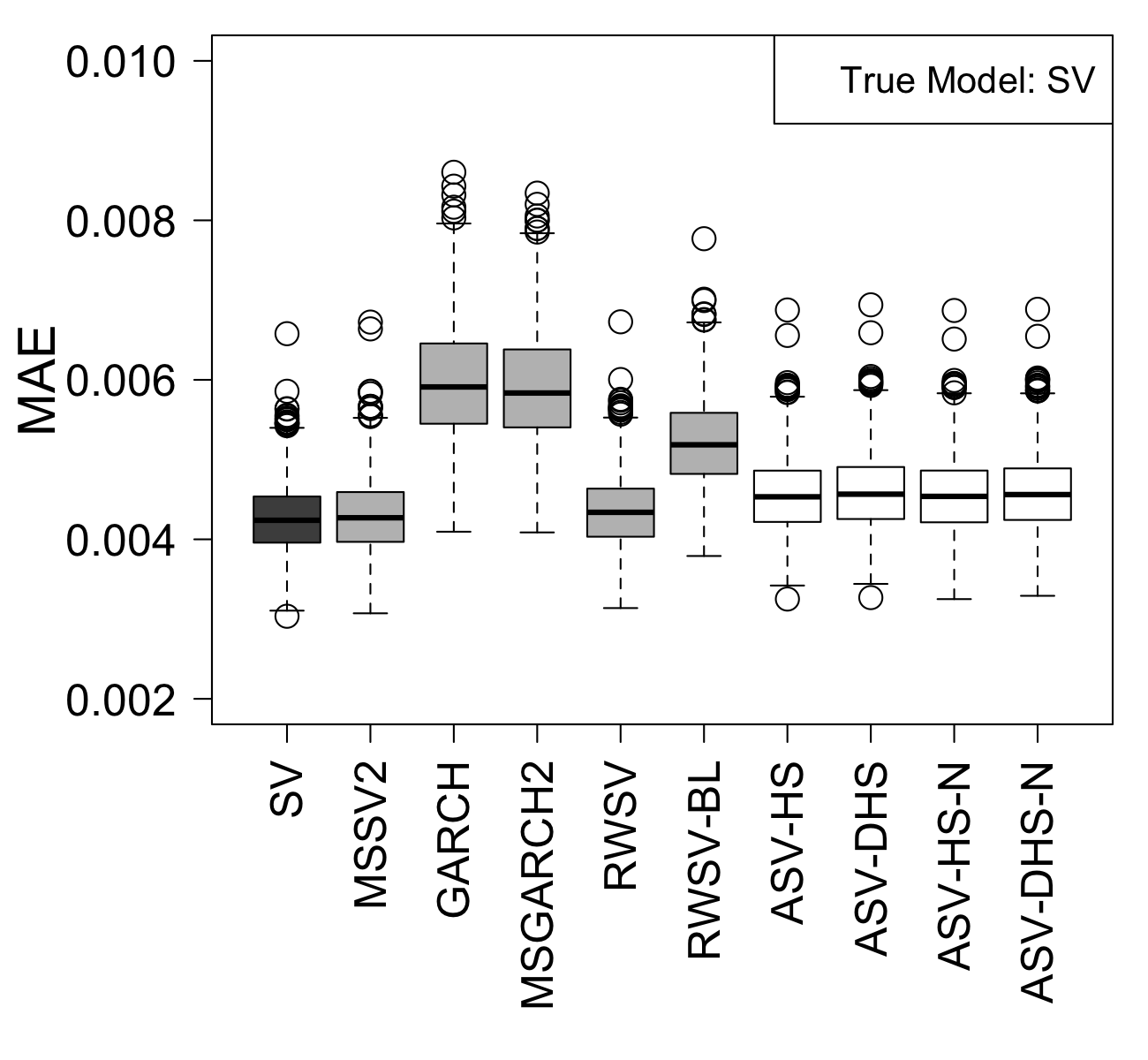}
      \caption{DGP 1}
      \label{fig:s1}
    \end{subfigure}
    \begin{subfigure}[b]{0.24\textwidth}
      \includegraphics[width=\linewidth]{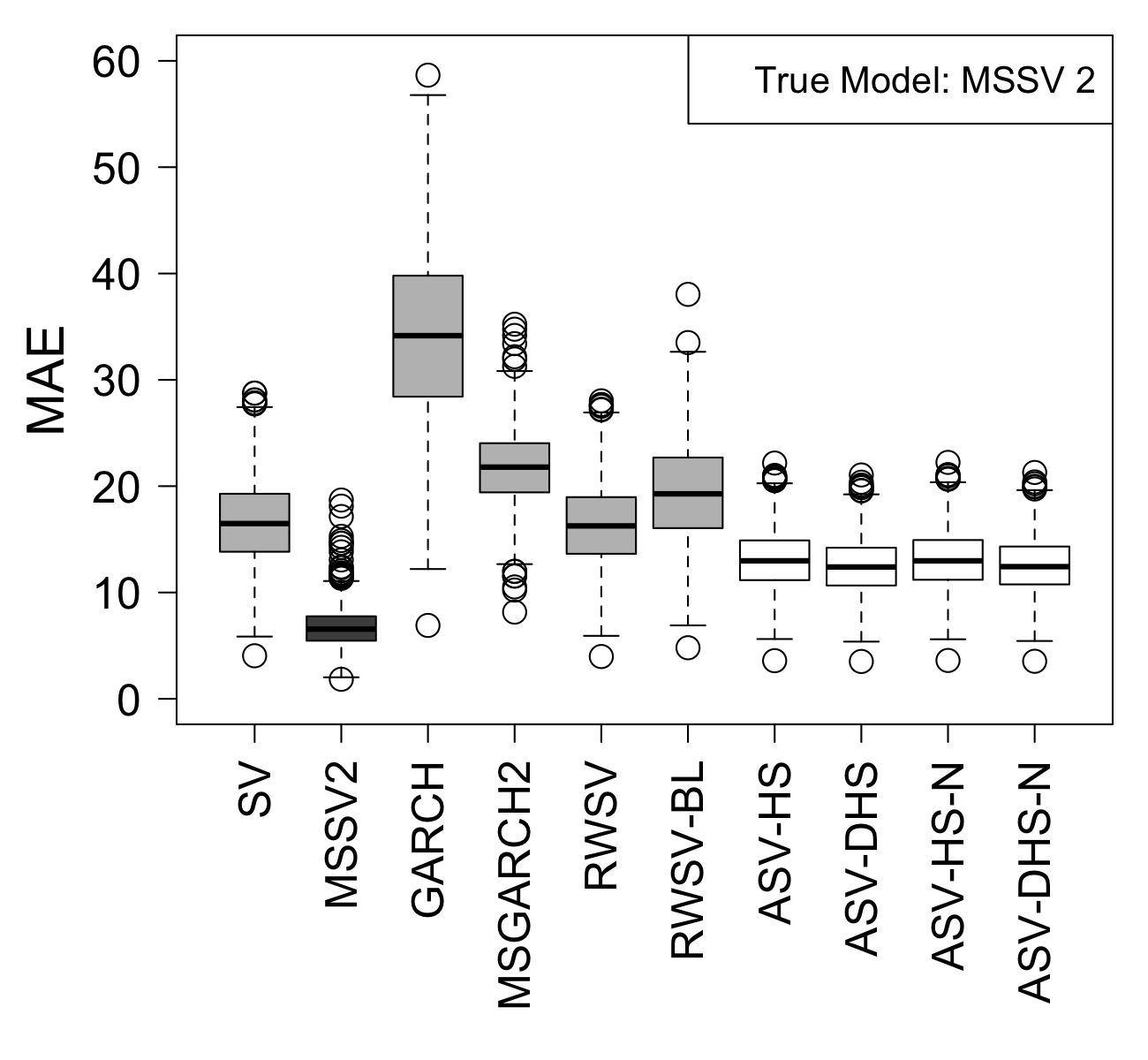}
      \caption{DGP 2}
      \label{fig:s2}
    \end{subfigure}
    \begin{subfigure}[b]{0.24\textwidth}
      \includegraphics[width=\linewidth]{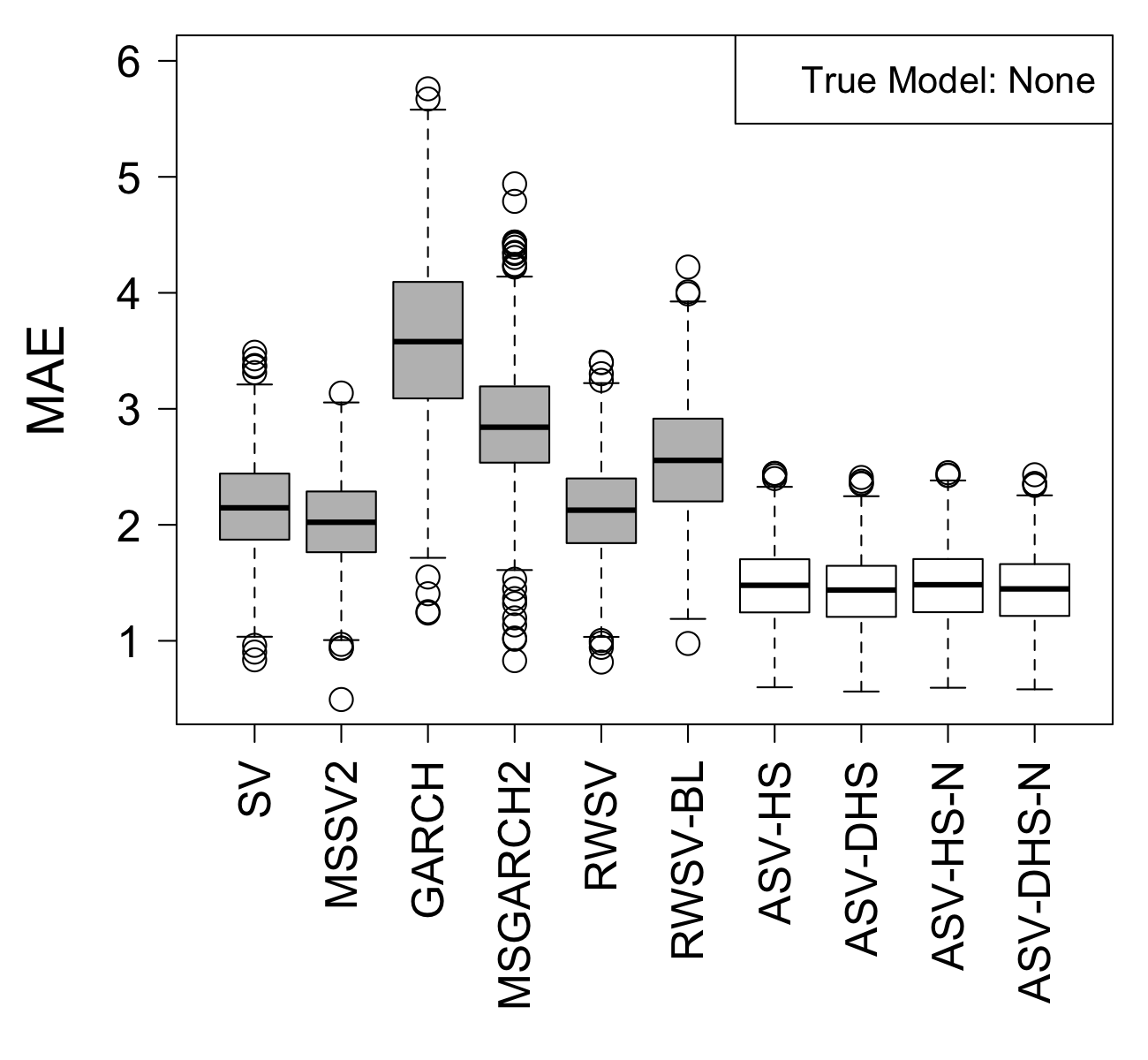}
      \caption{DGP 3}
      \label{fig:s3}
    \end{subfigure}
    \begin{subfigure}[b]{0.24\textwidth}
      \includegraphics[width=\linewidth]{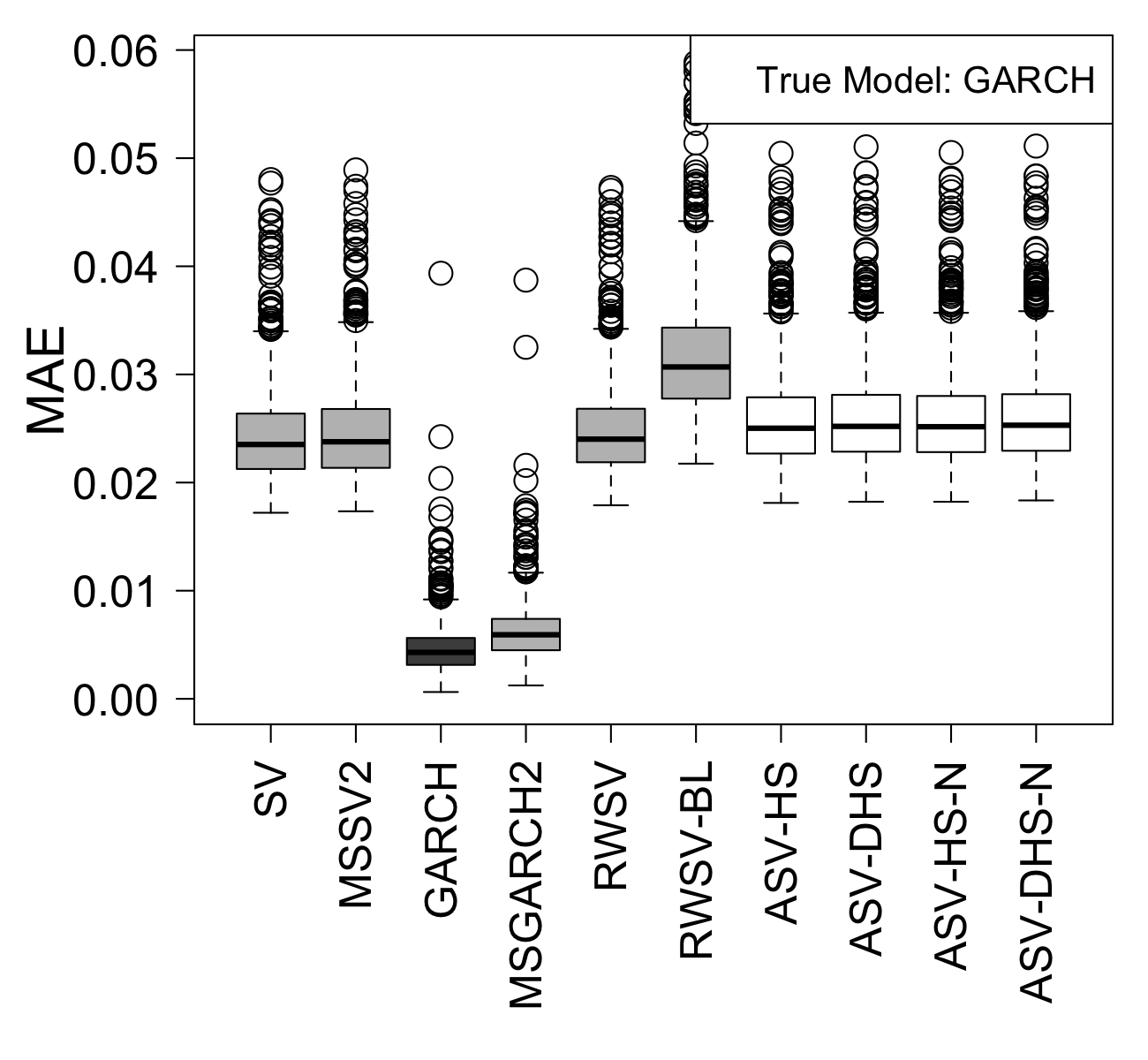}
      \caption{DGP 4}
      \label{fig:s4}
    \end{subfigure}
    \vspace{1em}
    \begin{subfigure}[b]{0.24\textwidth}
      \includegraphics[width=\linewidth]{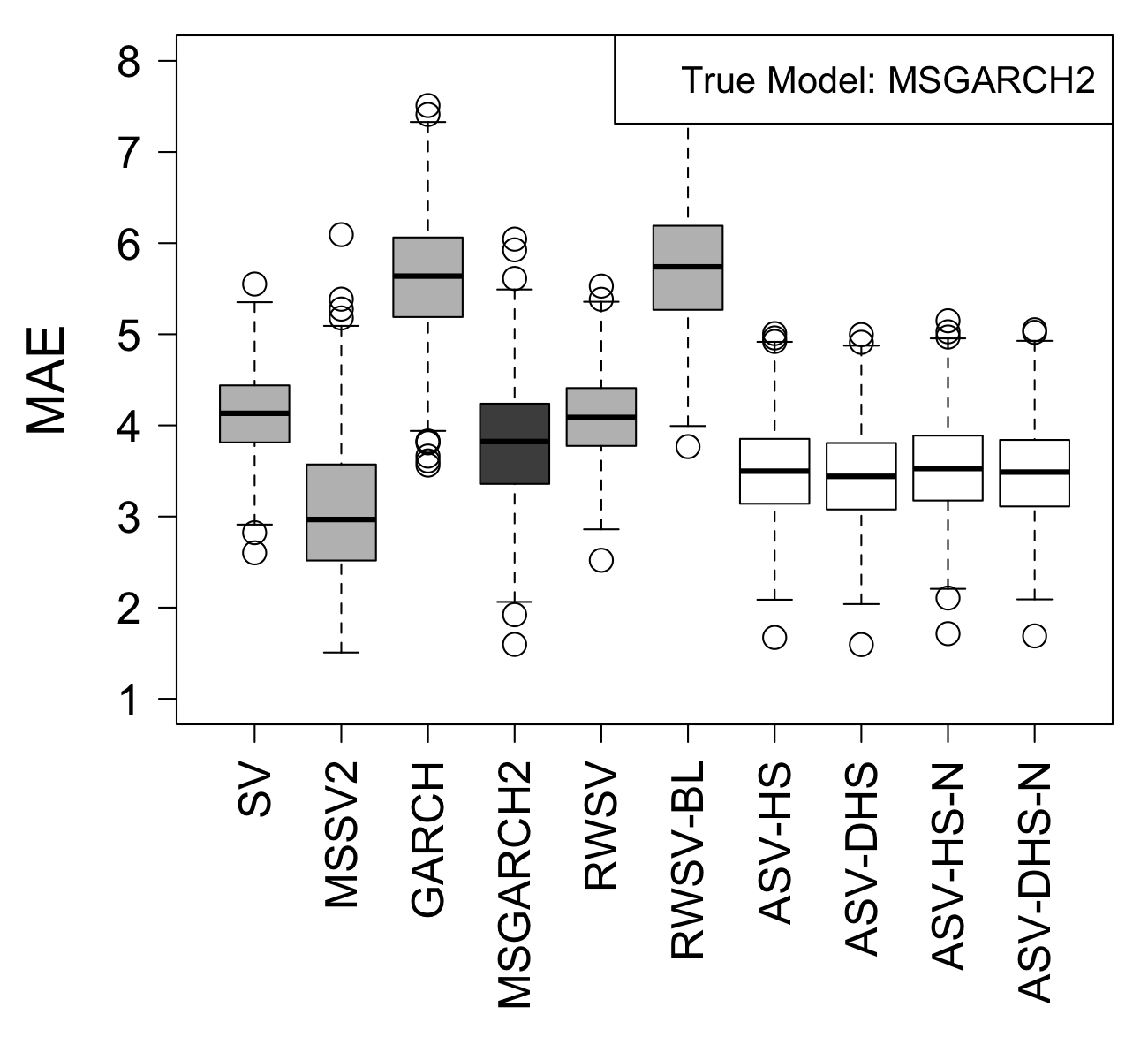}
      \caption{DGP 5}
      \label{fig:s5}
    \end{subfigure}
    \begin{subfigure}[b]{0.24\textwidth}
      \includegraphics[width=\linewidth]{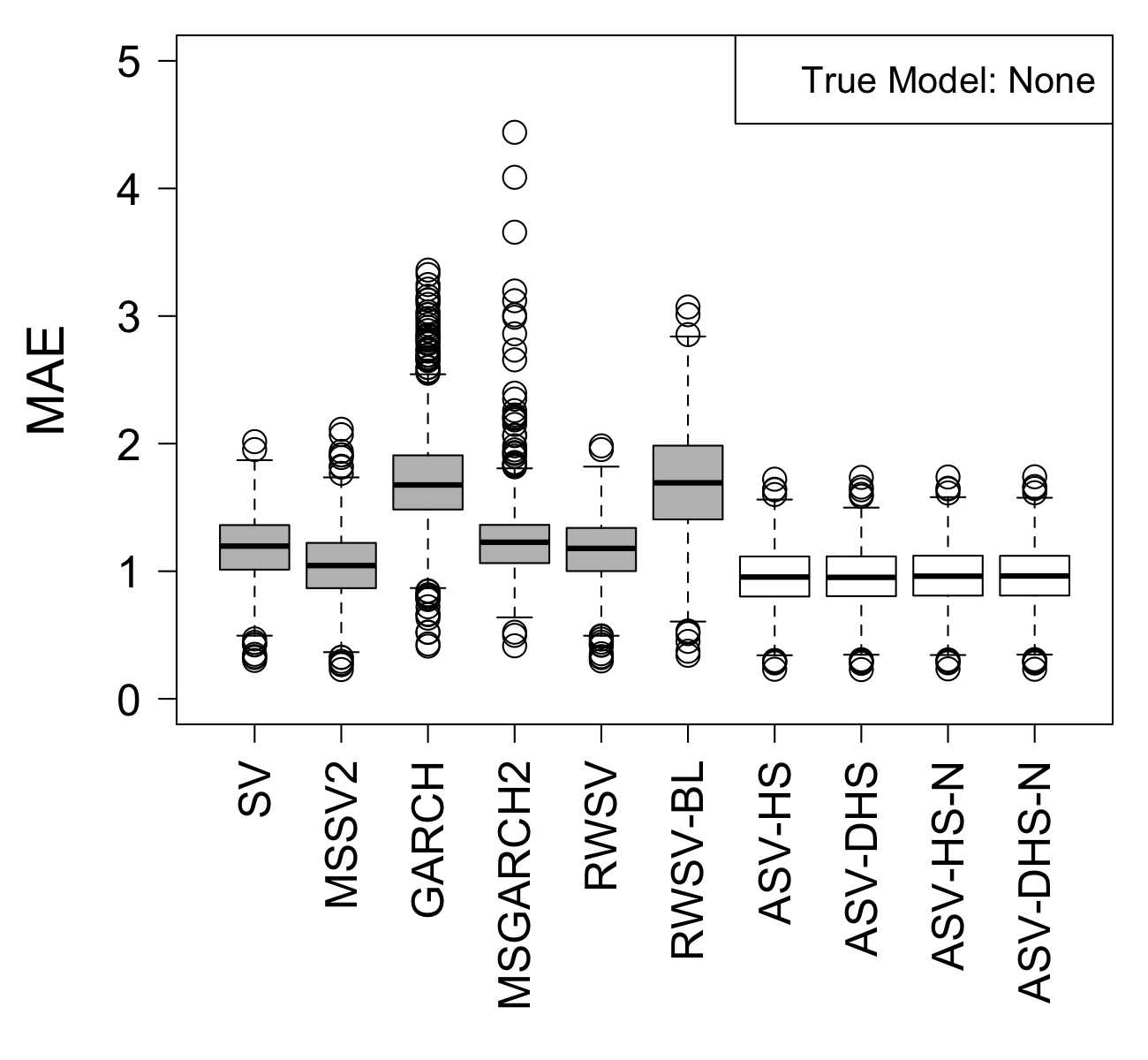}
      \caption{DGP 6}
      \label{fig:s6}
    \end{subfigure}
    \begin{subfigure}[b]{0.24\textwidth}
      \includegraphics[width=\linewidth]{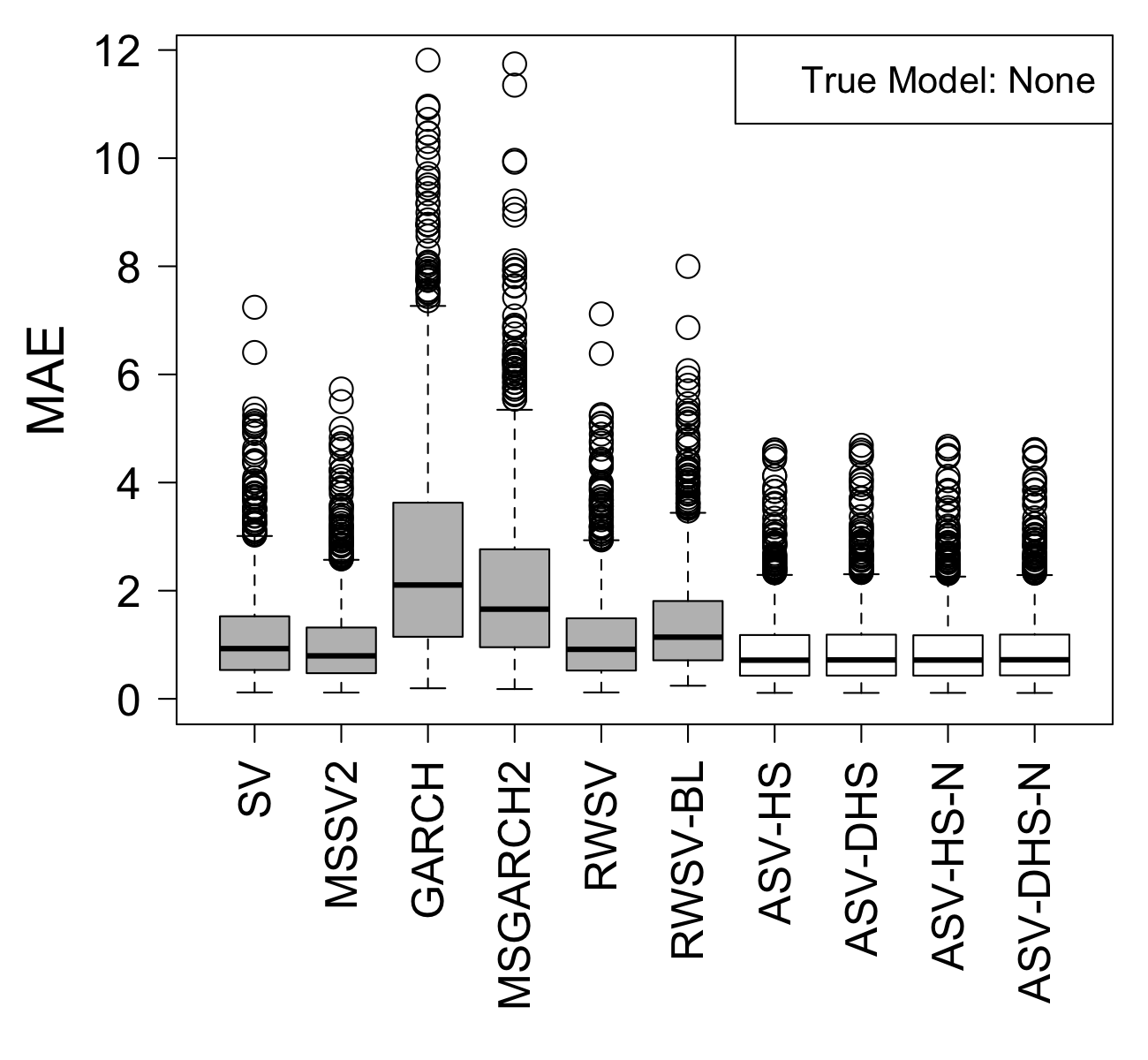}
      \caption{DGP 7}
      \label{fig:s7}
    \end{subfigure}
    \begin{subfigure}[b]{0.24\textwidth}
      \includegraphics[width=\linewidth]{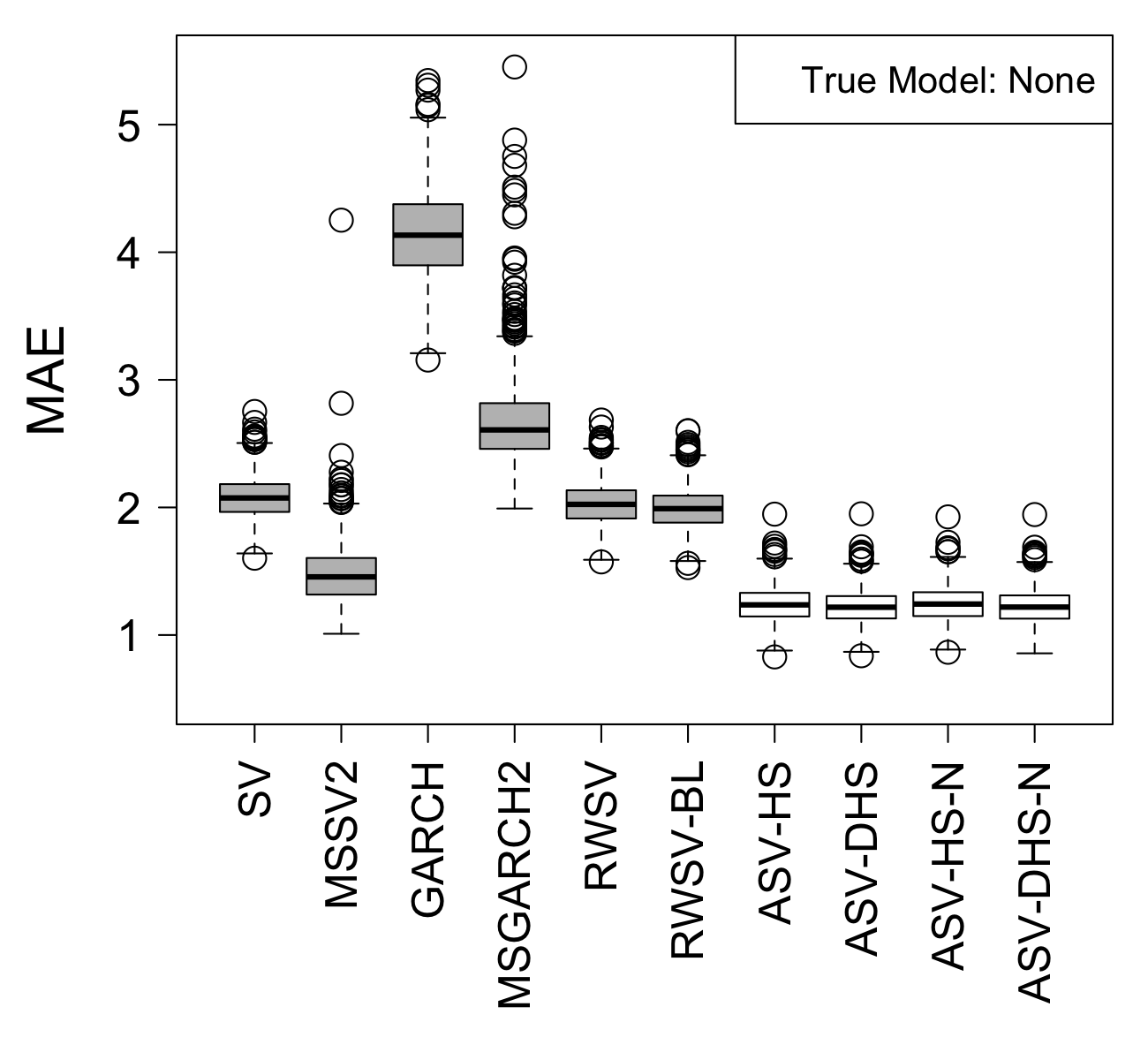}
      \caption{DGP 8}
      \label{fig:s8}
    \end{subfigure}
    \caption{Box plots of mean absolute error (MAE) across 1,000 sample paths comparing Stochastic Volatility (SV), Markov-Switching Stochastic Volatility with 2 Regimes (MSSV2), Generalized Autoregressive Conditional Heteroskedasticity (GARCH), Markov-Switching GARCH with 2 Regimes (MSGARCH2), Random Walk SV with Inverse-Gamma Prior (RWSV), Random Walk SV with Bayesian LASSO (RWSV-BL), Adaptive Stochastic Volatility with the Horseshoe prior (ASV-HS), and Adaptive Stochastic Volatility with the Dynamic Horseshoe prior (ASV-DHS). Perfectly specified models, drawn in black, are SV and MSSV2 for DGPs 1 and 2, and GARCH and MSGARCH2 for DGPs 4 and 5. All models are misspecified for DGPs 3,6,7 and 8.
    }
    \label{fig:MAE_boxplots}
\end{sidewaysfigure}
As shown in Table~\ref{tab:wide_results} and Figure~\ref{fig:MAE_boxplots}, the proposed ASV models consistently outperform or match the true models in DGPs with large volatility shifts, including DGPs 2, 3, 5, 6, and 8. In DGPs 2 and 5, MSSV2 achieves the lowest MAE, as expected, since it assumes two distinct volatility states that match the structure of these DGPs. The proposed ASV models, despite not assuming a fixed two-state structure, consistently achieve the second-lowest MAE in both cases. 

ASV models outperform existing alternatives particularly when large and abrupt changes in volatility are present. In DGP 3, the average MAE of the ASV models is 1.44 (sd = 0.32), compared with 2.03 (sd = 0.38) for the next best model, MSSV2, corresponding to a 29\% reduction. Similarly, DGP 8 shows an 18\% reduction in MAE, and DGP 6 shows a 9\% reduction relative to the second-best model. Despite having three distinct states, existing models perform relatively well in DGP 6. This is primarily due to the high persistence parameter, $\beta$, characterizing DGP 6, which enables autoregressive models to track volatility changes more effectively. When persistence is lower, as in DGPs 3 and 8, the advantage of ASV becomes substantially more pronounced. These results highlight the strong ability of ASV to capture abrupt structural shifts in volatility dynamics.

ASV models also perform reasonably well in scenarios where the underlying volatility is stationary or smoothly varying, as in DGPs 1 and 7. In DGP 1, the SV model achieves the lowest average MAEs, as it is the correctly specified model. The ASV models perform similarly to other misspecified SV type models, such as RWSV and MSSV2. In DGP 7, where volatility evolves smoothly without abrupt shifts, the ASV models outperform all alternatives, including SV and RWSV. This suggests that although ASV is designed to capture abrupt structural changes, it remains competitive in settings characterized by smooth or mean-reverting dynamics. ASV performs relatively poorly in DGP 4, showing MAE values comparable to other SV-type models. The reduced performance likely stems from model misspecification, as DGP 4 follows a GARCH(1,1) process while ASV assumes an SV formulation. GARCH volatility is a deterministic function of past data and therefore produces a much rougher stationary volatility path than the smoother dynamics in SV models. This mismatch makes accurate volatility estimation more difficult for ASV.

The simulation study also highlights the limitations of regime-switching models such as MSSV2 and MSGARCH2, which rely on a fixed number of latent regimes. These models perform poorly when the assumed regime structure does not align with the true DGPs. This mismatch is evident in DGPs 3, 6, and 8, where the number of volatility levels exceeds the models' assumed structure. In contrast, the ASV models estimate volatility adaptively through a global-local shrinkage prior, allowing flexible adjustment to a wide range of structural patterns without requiring a fixed regime specification.

The second and third rows of Table~\ref{tab:wide_results} present the ECs and MCIWs of the 90\% credible intervals for the eight Bayesian methods in the simulation study: SV, MSSV2, RWSV, RWSV-BL and four variants of ASV. They are not reported for GARCH and MSGARCH2, as these frequentist models yield deterministic estimates of conditional volatility given the fitted parameters and data. ASV-HS and ASV-DHS achieve near-nominal 90\% coverage in all DGPs except DGPs 1 and 4. In DGP 1, they exhibit substantial undercoverage, with empirical coverage dropping to 76\%, and in DGP 4, they show mild undercoverage at 84\%. This is likely due to the models' assumption of nonstationarity, which conflicts with the stationary structure of the DGPs 1 and 4. The nugget-augmented variants, ASV-HS-N and ASV-DHS-N, address this issue by introducing an additional noise term, achieving improved coverage of 86\% in DGP 1 and 87\% in DGP 4.

Across all DGPs, the ASV models produce narrow credible intervals, achieving the smallest MCIWs in seven of the eight settings. MSSV2 yields the narrowest intervals in DGP 2, where it is the correctly specified model. However, the undercoverage observed for ASV in DGP 1 suggests that its intervals are too narrow, leading to an underestimation of uncertainty. The ASV-N variants yield intervals that are only slightly wider than those of the base ASV models, yet remain generally narrower than those of other competing models. These results indicate that ASV models with a nugget term provide both precise and well-calibrated uncertainty quantification.

The main structural difference between the HS and DHS variants of ASV is the inclusion of the autoregressive shrinkage parameter, $\phi$, in ASV-DHS, which controls the temporal dependence of the local scale parameters. Compared with ASV-HS, ASV-DHS shows consistent but modest improvements in DGPs 2, 3, 5, and 8, where large and recurrent volatility shifts occur, with reductions in MAE of 5\%, 3\%, 2\%, and 2\%, respectively. These results suggest that dynamic shrinkage offers improved local adaptivity when volatility shifts are both abrupt and recurrent.
\section{Empirical Study}\label{sec:empiricalstudy}
\begin{sidewaysfigure}[!htbp]
    \centering
    \begin{subfigure}[b]{0.24\textwidth}
      \includegraphics[width=\linewidth]{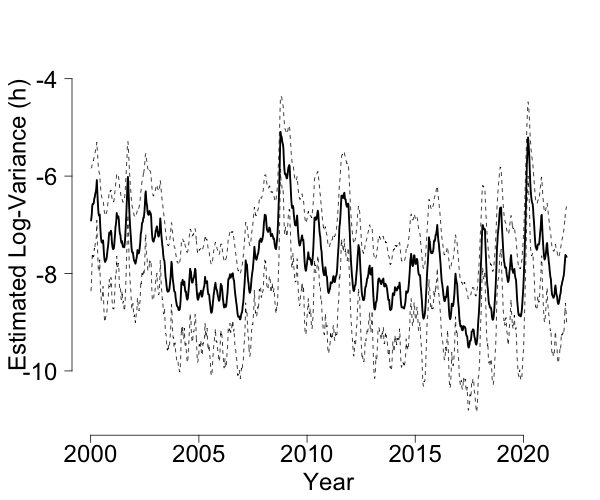}
      \caption{S\&P 500: SV}
      \label{fig:snp_sv}
    \end{subfigure}
    \begin{subfigure}[b]{0.24\textwidth}
      \includegraphics[width=\linewidth]{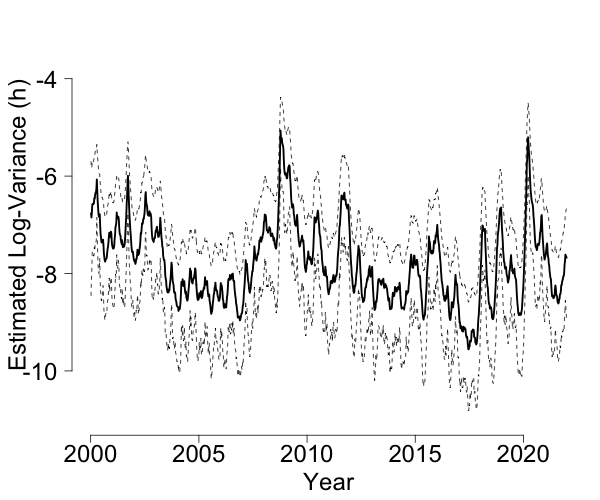}
      \caption{S\&P 500: MSSV2}
      \label{fig:snp_mssv}
    \end{subfigure}
    \begin{subfigure}[b]{0.24\textwidth}
      \includegraphics[width=\linewidth]{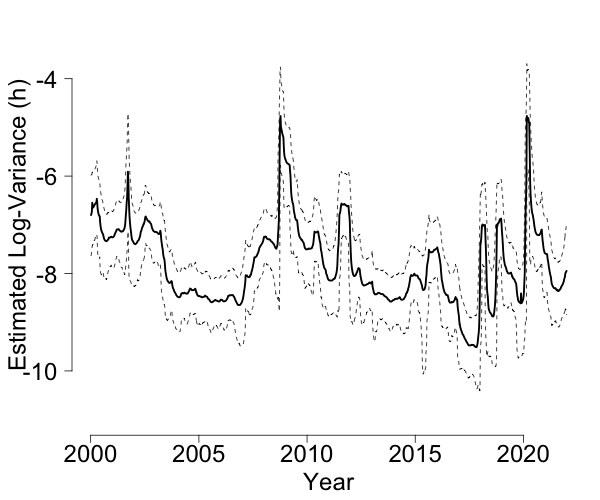}
      \caption{S\&P 500: ASV-DHS}
      \label{fig:snp_hs}
    \end{subfigure}
    \begin{subfigure}[b]{0.24\textwidth}
      \includegraphics[width=\linewidth]{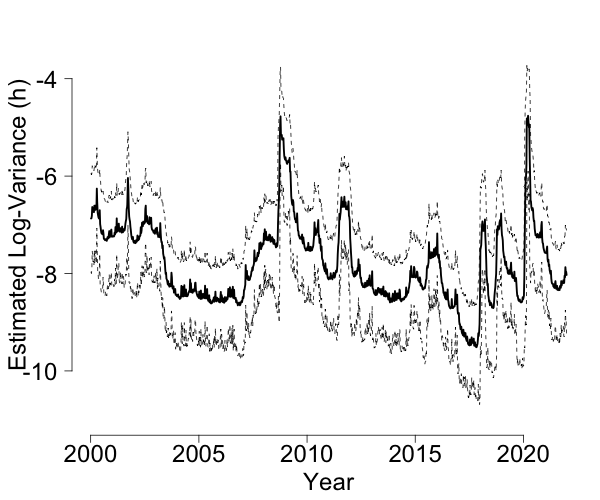}
      \caption{S\&P 500: ASV-DHS-N}
      \label{fig:snp_hsn}
    \end{subfigure}

    \begin{subfigure}[b]{0.24\textwidth}
      \includegraphics[width=\linewidth]{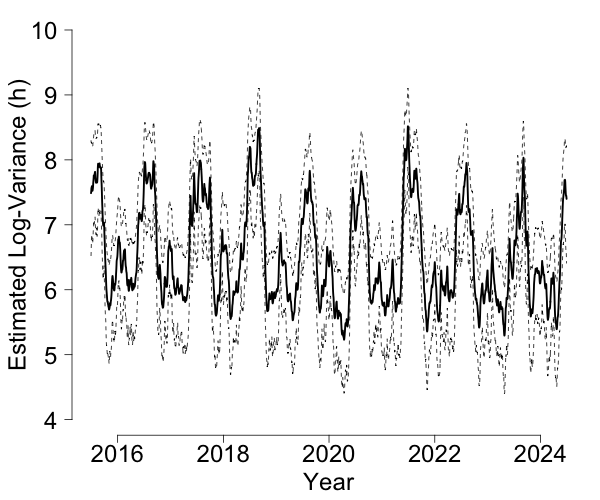}
      \caption{Electricity: SV}
      \label{fig:elec_sv}
    \end{subfigure}
    \begin{subfigure}[b]{0.24\textwidth}
      \includegraphics[width=\linewidth]{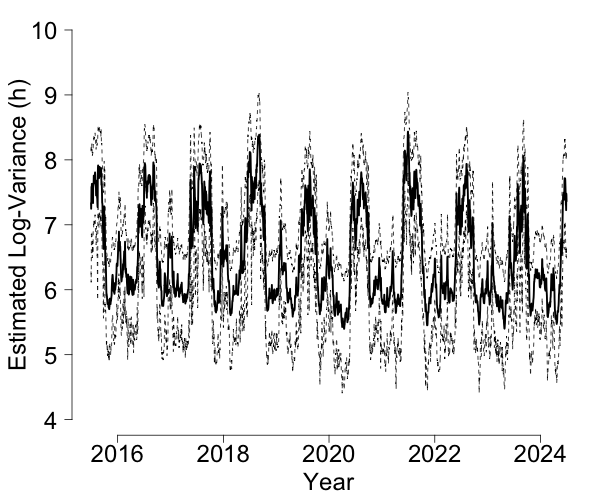}
      \caption{Electricity: MSSV2}
      \label{fig:elec_mssv}
    \end{subfigure}
    \begin{subfigure}[b]{0.24\textwidth}
      \includegraphics[width=\linewidth]{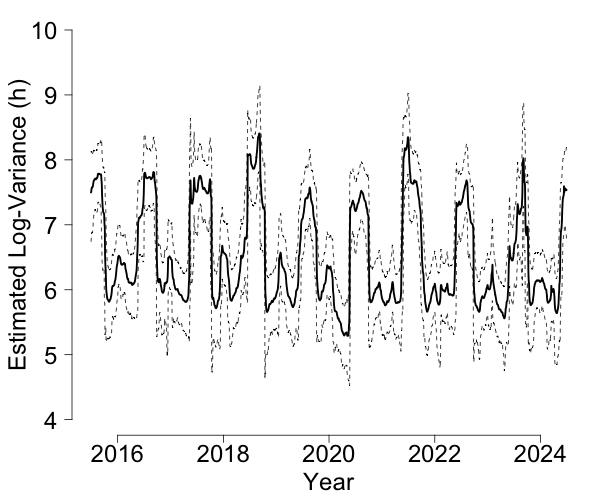}
      \caption{Electricity: ASV-DHS}
      \label{fig:elec_hs}
    \end{subfigure}
    \begin{subfigure}[b]{0.24\textwidth}
      \includegraphics[width=\linewidth]{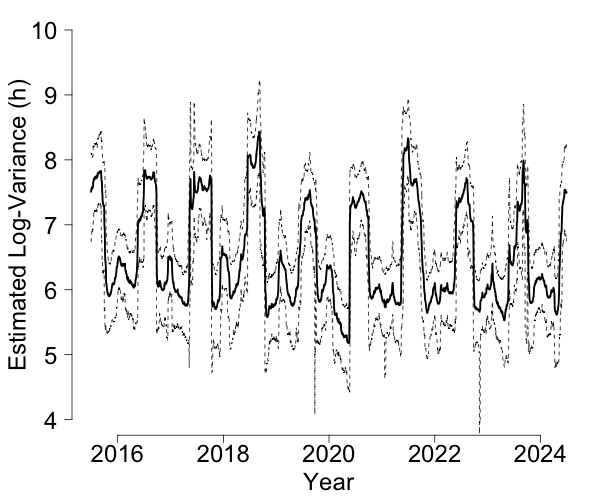}
      \caption{Electricity: ASV-DHS-N}
      \label{fig:elec_hsn}
    \end{subfigure}

    \begin{subfigure}[b]{0.24\textwidth}
      \includegraphics[width=\linewidth]{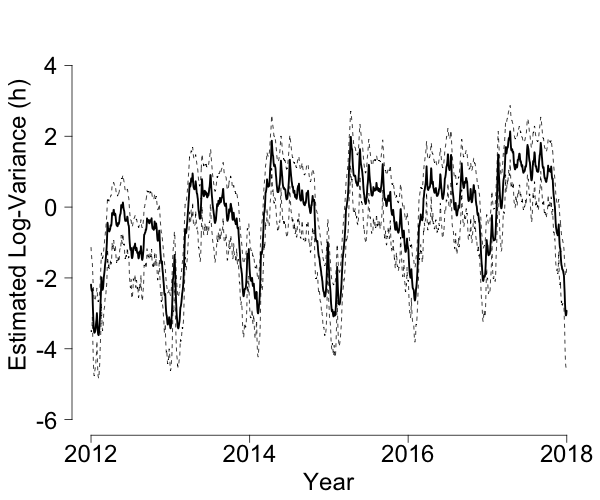}
      \caption{Bike Rental: SV}
      \label{fig:bike_sv}
    \end{subfigure}
    \begin{subfigure}[b]{0.24\textwidth}
      \includegraphics[width=\linewidth]{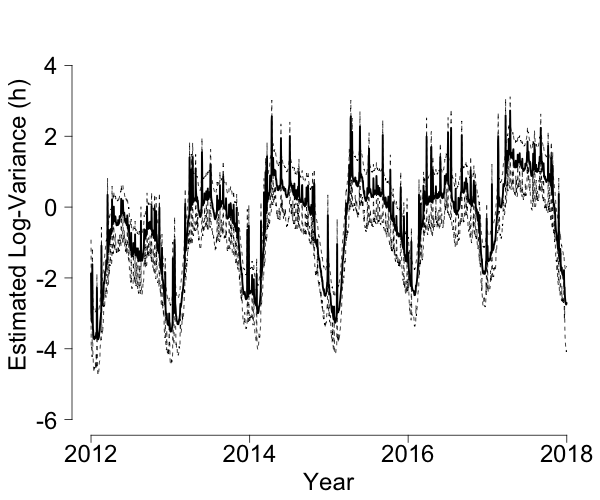}
      \caption{Bike Rental: MSSV2}
      \label{fig:bike_mssv}
    \end{subfigure}
    \begin{subfigure}[b]{0.24\textwidth}
      \includegraphics[width=\linewidth]{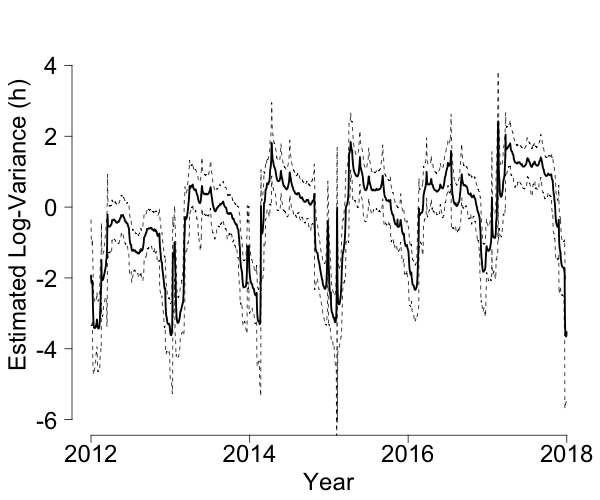}
      \caption{Bike Rental: ASV-DHS}
      \label{fig:bike_hs}
    \end{subfigure}
    \begin{subfigure}[b]{0.24\textwidth}
      \includegraphics[width=\linewidth]{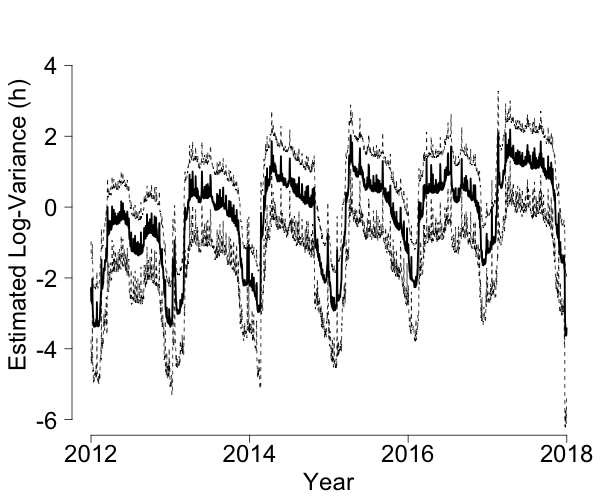}
      \caption{Bike Rental: ASV-DHS-N}
      \label{fig:bike_hsn}
    \end{subfigure}
    \caption{Estimated log-variance from four methods, Stochastic Volatility (SV), Markov-Switching SV with two regimes (MSSV2), Adaptive Stochastic Volatility with the dynamic horseshoe prior (ASV-DHS), and ASV-DHS with a nugget effect, applied to three datasets: weekly S\&P 500 log returns, daily electricity demand, and daily bike rental counts.}
    \label{fig:emp_data}
\end{sidewaysfigure}
We continue our analysis of the datasets introduced in Section~\ref{sec:intro} and estimate the log-variance using four methods: SV and MSSV2, which serve as established baselines, and ASV-DHS and ASV-DHS-N, our proposed models. For the ASV models, we set $k = 1$ to reflect the presence of abrupt changes observed in Figure~\ref{fig:Motivation}. In the S\&P 500 dataset (\Cref{fig:snp_sv,fig:snp_mssv,fig:snp_hs,fig:snp_hsn}), all models detect major volatility spikes in 2009 and 2020, corresponding to the global financial crisis and the COVID-19 pandemic. Moderate increases are also observed around 2001 (9/11), 2011 (European debt crisis), and 2018 (U.S. government shutdowns). ASV-DHS and ASV-DHS-N provide smoother and more stable estimates compared to SV and MSSV2, particularly during low-volatility periods. For both the electricity (\Cref{fig:elec_sv,fig:elec_mssv,fig:elec_hs,fig:elec_hsn}) and bike rental datasets (\Cref{fig:bike_sv,fig:bike_mssv,fig:bike_hs,fig:bike_hsn}), all methods capture strong seasonal patterns in volatility, with elevated uncertainty coinciding with periods of peak activity. In the electricity dataset, this corresponds to high-demand months, while in the bike rental dataset, volatility increases during the summer.
\begin{table}[ht]
    \centering
    \begin{tabular}{llcccc}
    \toprule
    \textbf{Dataset} & \textbf{Metric} & SV & MSSV2 & ASV-DHS & ASV-DHS-N \\
    \midrule
    \multirow{3}{*}{S\&P 500} 
    & $\mathrm{Mean}$ & 0.0739 & 0.0792 & 0.037 & 0.0799 \\
    & $\mathrm{Kurtosis}$ & 0.9105 & 0.7348 & 52.3562 & 11.5863 \\
    & $\mathrm{CP}$ & 0 & 0 & 9 & 3 \\
    & $\mathrm{LL}$ & 1.784 & 1.782 & 1.803 & 1.789\\
    \midrule
    \multirow{3}{*}{Electricity} 
    & $\mathrm{Mean}$ & 0.0307 & 0.0600 & 0.0176 & 0.0213 \\
    & $\mathrm{Kurtosis}$ & 0.1691 & 7.5788 & 65.4116 & 68.4159 \\
    & $\mathrm{CP}$ & 0 & 4 & 16 & 20 \\
    & $\mathrm{LL}$ & 1.831   & 1.862  & 1.815 & 1.816 \\
    \midrule
    \multirow{3}{*}{Bike} 
    & $\mathrm{Mean}$ & 0.0616 & 0.143 & 0.0456 & 0.1331 \\
    & $\mathrm{Kurtosis}$ & 0.1989 & 13.3789 & 90.7971 & 5.6332 \\
    & $\mathrm{CP}$ & 0 & 3 & 10 & 2 \\
    & $\mathrm{LL}$ & 2.708 & 2.710 & 2.648 & 2.564\\
    \bottomrule
    \end{tabular}
    \caption{Empirical results: (1) the mean and (2) kurtosis of the estimated log-variance increments $|\Delta \hat{h}_t|$ and (3) the total number of flagged change points (CP) and (4) Logarithmic Loss (LL) on short term forecasts (6 step-ahead).  CP are computed as $\sum_{t=2}^{T} \mathds{1}\{|\Delta \hat{h}_t| > \delta \}$, where $\delta := 5 \times \texttt{sd}(\Delta \hat{h}_t)$. LL is computed as $\frac{1}{6}\sum_{t=1}^{6}\left|\log(y_{T+t}^2) - \hat{h}_{T+t}\right|$, where $T$ denotes the length of the training sample and $\hat{h}_{t}$ is the estimated log-variance. Each model is first estimated using the first 100 observations, then re-estimated using the first 200, 300, and so on, with a 6-step-ahead forecast generated at each stage until the end of the sample.}\label{tab:summary_stats}
\end{table}

Table~\ref{tab:summary_stats} presents summary metrics of the estimated log-variance increments $|\Delta \hat{h}_{t}|$ across the three empirical datasets. These metrics, mean, kurtosis, the number of flagged change points (CP), and the logarithmic loss (LL) computed from short-term (6-step-ahead) forecasts. The first three metrics offer insight into the smoothness, tail behavior, and local adaptivity, while LL offers a rough indication of short-term predictive accuracy \citep{pagan}. 

ASV-DHS consistently produces the lowest mean, indicating that, on average, it exhibits the smallest changes in the estimated volatility compared to other models. At the same time, it shows extremely high kurtosis and the highest number of flagged change points, suggesting heavy-tailed behavior in the increments and strong local adaptivity to sudden structural shifts. These properties highlight the model's tendency to promote overall smoothness while remaining responsive to large, isolated changes.

The smoothness and local adaptability of ASV-DHS-N vary depending on the underlying structure of the data. In the electricity dataset, where volatility is characterized by extended periods of small changes punctuated by occasional abrupt shifts, the mean in Table~\ref{tab:summary_stats} under ASV-DHS-N is similar to that of ASV-DHS, indicating comparable overall variation. In contrast, for the bike rental and S\&P 500 datasets, which exhibit more unpredictable changes in volatility, the mean and CP patterns resemble those of MSSV2, suggesting behavior similar to a smoother regime-switching model. The kurtosis remains substantially higher than that of SV and MSSV2, preserving the model's ability to capture heavy-tailed behavior. 

The LL results indicate that the proposed models, ASV-DHS and ASV-DHS-N, despite not being developed as forecasting models, perform comparably to benchmark approaches overall, with slightly lower LL for the electricity and bike rental datasets. The notably lower LL values for the electricity and bike rental data are consistent with their clear nonstationary volatility patterns, characterized by abrupt and irregular changes. In contrast, the smaller differences among models for the S\&P 500 data suggest comparatively more stable volatility dynamics over the sample period.
\begin{figure}[ht]
    \centering 
    \begin{subfigure}[b]{0.32\textwidth}
      \includegraphics[width=\linewidth]{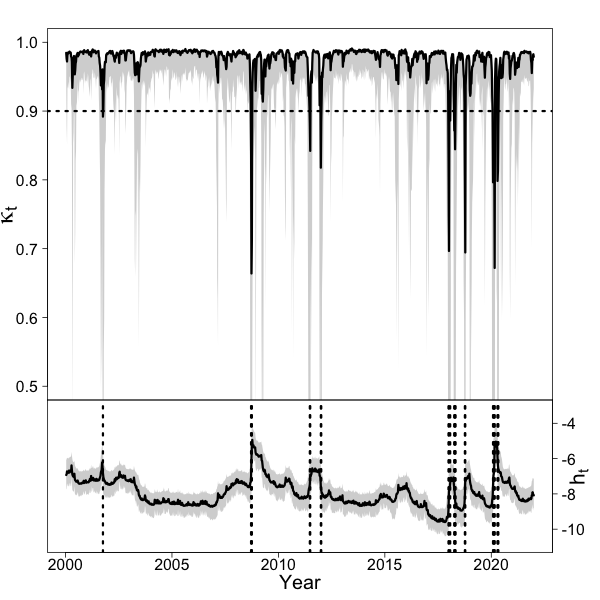}
      \caption{S\&P 500 Index}
      \label{fig:kappa_a}
    \end{subfigure}
        \begin{subfigure}[b]{0.32\textwidth}
      \includegraphics[width=\linewidth]{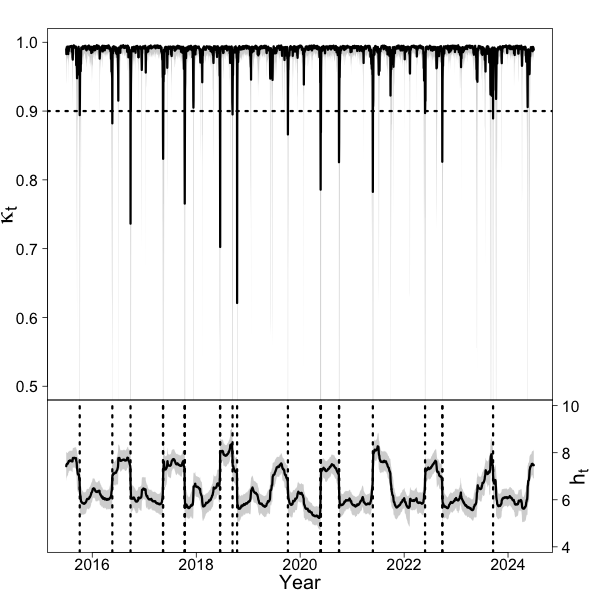}
      \caption{Electricity}
      \label{fig:kappa_b}
    \end{subfigure}
    \begin{subfigure}[b]{0.32\textwidth}
      \includegraphics[width=\linewidth]{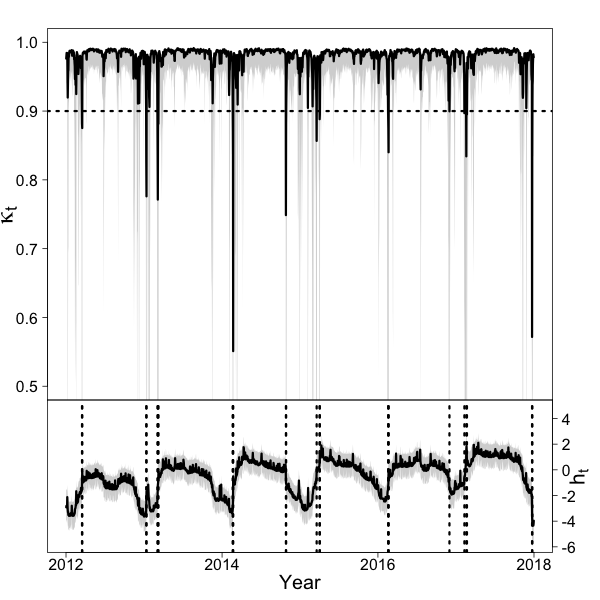}
      \caption{Bike Rentals}
      \label{fig:kappa_c}
    \end{subfigure}
\caption{Comparison of the posterior means of the shrinkage parameter, $\kappa_t := 1/(1+\exp{v_t})$, where $v_{t} = \log(\mathrm{Var}(\Delta h_t))$, and the log-variance parameter, $h_t$, estimated using ASV-DHS-N for the S\&P 500, electricity demand, and bike rental datasets. The shaded regions represent the 95\% and centered 90\% credible intervals for $\kappa_t$ and $h_t$, respectively. Vertical lines indicate the dates at which the expected $\kappa_t$ falls below 0.9.}\label{fig:kappa}
\end{figure}

Another defining feature of ASV is its ability to localize the timing of large variations in volatility through the shrinkage parameter $\kappa_t := \frac{1}{1 + \exp(v_t)}$. $\kappa_t$ close to 0 corresponds to high variability in the latent volatility increments $\Delta h_t$, indicating periods of substantial changes in $h_t$, while a $\kappa_t$ near 1 suggests strong shrinkage and minimal change. Thus $\kappa_t$ serves as an interpretable signal for identifying the timing of structural shifts in volatility.

In Figure~\ref{fig:kappa}, we highlight the time points in which the posterior mean of $\kappa_t$ falls below 0.9, marking potential periods of abrupt volatility changes. For the S\&P 500 index (\Cref{fig:kappa_a}), this includes notable episodes explored earlier, such as the 2008 financial crisis, the 2011 European debt crisis, and the COVID-19 shock in early 2020. In \Cref{fig:kappa_b}, volatility in electricity usage spikes around late spring and early fall, periods that typically mark transitions in heating or cooling demand, reflecting changing usage patterns driven by weather fluctuations. \Cref{fig:kappa_c} shows a similar pattern with volatility spikes in bike rental counts occurring near the start of spring, such as late February and March, and near the end-of-year holiday season in December and January.

\section{Trend Filtering Jointly in Mean and Variance}\label{sec:DSPmv}
The Bayesian Trend Filter with DSP (BTF-DSP) by \citet{dsp} provides a smooth and locally adaptive estimate of the mean of a time series. As an extension, we propose the Bayesian Trend Filter with ASV (BTF-ASV), thereby allowing locally adaptive estimates of both the mean $\{\beta_t\}_{t=1}^{T}$, and the log-variance process $\{h_{t}\}_{t=1}^{T}$ simultaneously:
\begin{equation}
    \begin{aligned}
	&y_{t} = \beta_{t} + \exp\{h_{t}/2\}\epsilon_{t}  && [\epsilon_{t}] \stackrel{iid}{\sim} N(0,1), \\ 
	& [\Delta^{k_\beta} \beta_{t}|\sigma^2_{\beta,t}] \sim N(0,\sigma^2_{\beta,t}) && [\log(\sigma^2_{\beta,t})|\mu_{\beta},\phi_{\beta}] \sim DSP(a_{\beta},b_{\beta},\mu_{\beta},\phi_{\beta}),\\ 
	& [\Delta^{k_h} h_{t}|\sigma^2_{h,t}] \sim N(0,\sigma^2_{h,t}) && [\log(\sigma^2_{h,t})|\mu_h,\phi_{h}] \sim DSP(a_{h},b_h,\mu_h,\phi_h). \\ 
    \end{aligned}\label{eq:Model2}
\end{equation}
We impose two conditionally independent priors on the $k$th-order differences of $\beta_t$ and $h_t$. The model distinguishes changes in the mean from changes in volatility through its structural formulation: $\beta_t$ enters the likelihood additively, while $h_t$ governs the multiplicative noise scale. This separation ensures that fluctuations in the variance do not influence the location of the distribution. Below, we show that the model is identifiable in the sense that the evolution variance terms, $\sigma^2_{\beta,t}$ and $\sigma^2_{h,t}$, are uniquely determined by the observed data.
\begin{theorem}\label{te:4} The following model is identifiable:
    \begin{align*}
        y_{t} = \beta_t + \exp\{h_{t}/2\}\epsilon_t & \quad [\epsilon_{t}]\stackrel{iid}{\sim}N(0,1), \\
        [\Delta^{k} \beta_t|\sigma^2_{\beta,t}] \sim N(0,\sigma_{\beta,t}^2) & \quad [\Delta^{k} h_{t}|\sigma^2_{h,t}] \sim N(0,\sigma_{h,t}^2).
    \end{align*}
\end{theorem}
\begin{proof}
Let $\boldsymbol{\beta} := \{\beta_1, \ldots, \beta_T\}$ and $\boldsymbol{h} := \{h_1, \ldots, h_T\}$ denote the latent mean and log-volatility processes, respectively.  
For notational convenience, define $\boldsymbol{\lambda}_{\beta} := \{\sigma_{\beta,1}^2, \ldots, \sigma_{\beta,T}^2\}$ and $\boldsymbol{\lambda}_{h} := \{\sigma_{h,1}^2, \ldots, \sigma_{h,T}^2\}$. For contradiction, fix two distinct parameter vectors $\boldsymbol{\theta}: = (\boldsymbol{\beta}, \boldsymbol{h},\boldsymbol{\lambda}_{\beta},\boldsymbol{\lambda}_h)$ and $ \boldsymbol{\theta}^{*} := (\boldsymbol{\beta}^*, \boldsymbol{h}^*,\boldsymbol{\lambda}_{\beta}^*,\boldsymbol{\lambda}_h^*)$ such that $\boldsymbol{\theta} \neq \boldsymbol{\theta}^*$  and $f(y|\boldsymbol{\theta})=f(y|\boldsymbol{\theta}^{*})$. Indeed,  $\mathrm{Var}(y|\boldsymbol{\theta}) = \mathrm{Var}(y|\boldsymbol{\theta}^{*}).$ This covariance equality necessarily forces $\boldsymbol{\theta}=\boldsymbol{\theta}^{*}$, leading to contradiction. Detailed proof is shown in Appendix A.
\end{proof}

While Theorem~\ref{te:4} establishes the theoretical identifiability of the BTF-ASV model, practical identifiability, the extent to which the available data are informative enough to separate the first-order effect from the second-order effects, may still be limited under severe model misspecification or high noise in the analyzed data. Practical identifiability in Bayesian models is an active area of research \citep{pident1,pident3,pident2}. Extending such ideas to time-series models may be an interesting direction for future research.
\begin{figure}[ht]
    \centering 
    \begin{subfigure}[b]{0.32\textwidth}
      \includegraphics[width=\linewidth]{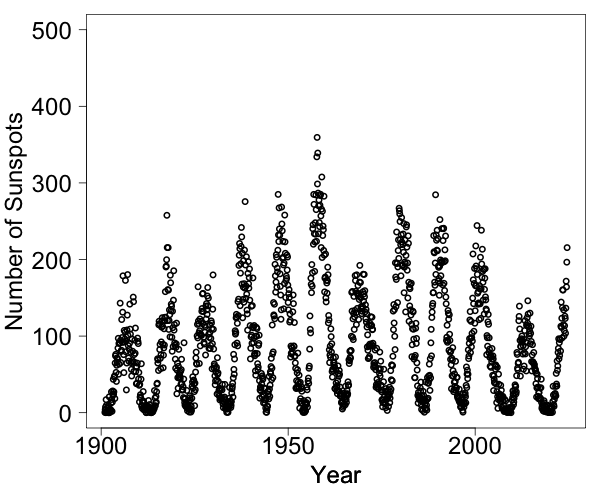}
      \caption{Observed Series $y_{t}$}
      \label{fig:btfasv_a}
    \end{subfigure}
        \begin{subfigure}[b]{0.32\textwidth}
      \includegraphics[width=\linewidth]{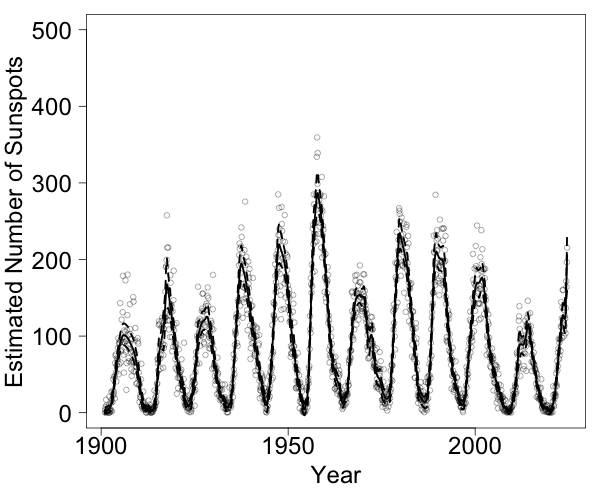}
      \caption{Estimated $\beta_{t}$ with 90\% CR}
      \label{fig:btfasv_b}
    \end{subfigure}
    \begin{subfigure}[b]{0.32\textwidth}
      \includegraphics[width=\linewidth]{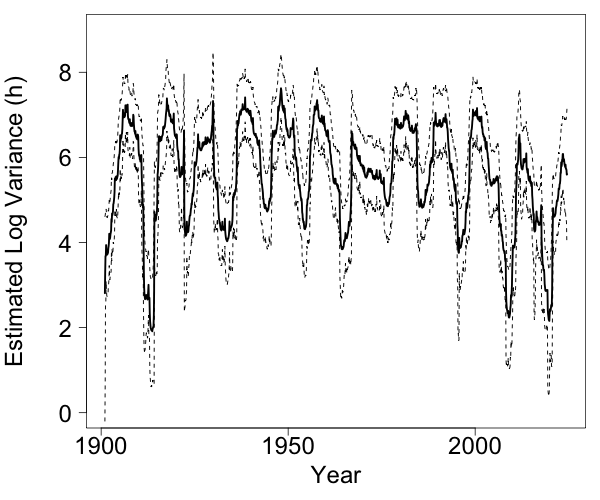}
      \caption{Estimated $h_t$ with 90\% CR}
      \label{fig:btfasv_c}
    \end{subfigure}
\caption{Monthly mean total sunspot numbers, defined as the average of daily total sunspot counts over each calendar month from January 1901 to August 2024, are shown in Figure~\ref{fig:btfasv_a}. The Bayesian Trend Filter with Adaptive Stochastic Volatility (BTF-ASV) is used to estimate the mean trend, $\beta_t$, as shown in Figure~\ref{fig:btfasv_b}. The estimated log-variance, $h_t$, along with its 90\% credible intervals, is presented in Figure~\ref{fig:btfasv_c}.}
\end{figure}

BTF-ASV is applied to the sunspot data to estimate both the time-varying mean and the volatility. The data consist of monthly mean total sunspot numbers, defined as the average daily total over each calendar month, from January 1901 to August 2024, obtained from \citet{sunspot}. The number of sunspots exhibits a well-known cyclic pattern associated with the 11-year solar cycle, which reflects periodic fluctuations in solar magnetic activity. This cyclical behavior is clearly visible in the raw observations shown in Figure~\ref{fig:btfasv_a}. The BTF-ASV estimate of the mean trend, $\beta_t$, shown in Figure~\ref{fig:btfasv_b}, provides a smooth reconstruction of this underlying cycle.

Interestingly, the estimated log-variance (Figure~\ref{fig:btfasv_c}) also follows a cyclical pattern that closely aligns with the mean trend. Periods of increased solar activity are associated with elevated volatility, suggesting that not only does solar activity rise and fall periodically, but the variability around these phases also intensifies. This highlights BTF-ASV's ability to decompose noisy data into structured patterns in both the mean and variance processes, offering a richer understanding of solar dynamics.

\section{Conclusion}
We propose ASV, a flexible and locally adaptive model for capturing both abrupt and gradual changes in volatility. ASV extends RWSV by incorporating a global-local shrinkage prior, specifically the DSP by \citet{dsp}. We develop and study multiple variants of ASV, including models with the horseshoe (ASV-HS), dynamic horseshoe (ASV-DHS), and their nugget-augmented counterparts (ASV-HS-N and ASV-DHS-N).

Through extensive simulation studies, we show that ASV-HS-N and ASV-DHS-N consistently achieve low MAE, correct nominal coverage, and narrower credible intervals compared to existing models. Empirical applications demonstrate that ASV models effectively capture large structural changes in volatility while preserving smooth estimates in stable periods. Moreover, the shrinkage parameter $\kappa_t$ provides a useful diagnostic for further identifying the timing of structural shifts.

Finally, we integrate ASV into the error process of the Bayesian Trend Filter proposed in \citet{dsp}, resulting in the BTF-ASV model. Empirical studies reveal its ability to produce smooth and locally adaptive estimates of both the time-varying mean and variance. Future research directions could explore the applications of ASV and BTF-ASV in diverse fields such as finance, environmental science, and epidemiology, leveraging their capacity to produce adaptive estimates for time-series that exhibit abrupt changes in both the mean and variance.
\bigskip
\if1\blind
{
\section{Acknowledgment}
Financial support from National Science Foundation grants OAC-1940124 and DMS-2114143 is gratefully acknowledged. 
} \fi
\if0\blind
{
\section{Acknowledgment}
Acknowledgments will be provided in the final version.
} \fi
\section{Disclosure Statement}
The authors report there are no competing interests to declare.

\setstretch{1}
\bibliographystyle{abbrvnat}
\bibliography{ref.bib}

\appendix
\setstretch{1.9}
\section{Proofs of Theorems}
\subsection{Theorem 1}\label{sec:proof_T2}
	This proof directly follows from Kolmogorov's three series theorem, which states the sufficient and necessary condition for almost sure convergence of infinite sum of random variables. Let $\phi \in (0,1)$. Due to symmetry of $\text{sech}$ function, we only need to prove the case when $0<\phi<1$, and the case where $-1<\phi<0$, directly follows. For the first condition, let $\epsilon > 0$, and $\ell$ is some positive integer or 0. 
	\begin{align*}
		\mathrm{P}(|z_{\ell}| \geq \epsilon) &= 2\int_{\epsilon}^{\infty} \frac{\text{exp}\{x/(2\phi^{\ell})\}}{\pi \phi^\ell (1+ \text{exp}\{x/\phi^{\ell} \}} dx\\
		&= \frac{2}{\pi\phi^{\ell}} \int_{\epsilon}^{\infty} \frac{\text{exp}\{x/(2\phi^{\ell})\}}{1+ \text{exp}\{x/\phi^{\ell}\}} dx &  u = \text{exp}\{x/(2\phi^{\ell})\} \\ 
		&= \frac{4}{\pi} \int_{\epsilon}^{\infty} \frac{1}{1+ u^2}du & 2\phi^{\ell} du = \text{exp}\{x/(2\phi^{\ell})\}dx \\ 
		&= \frac{4}{\pi} \text{arctan}(\text{exp}\{x/(2\phi^{\ell})\})\bigg]_{\epsilon}^{\infty}\\
		&= 2(1-\frac{2}{\pi} \text{arctan}(\text{exp}\{\epsilon/(2\phi^{\ell})\}) \\
		&= \frac{4}{\pi} \text{arccotan}(\text{exp}\{\epsilon/(2\phi^{\ell})\} \\
		&\leq \frac{4}{\pi} \text{arccotan}(\text{exp}\{\ell\}) & \textit{if } \epsilon \leq 2 \ell \phi^\ell
	\end{align*}

Since $0<\phi<1$, $2 \ell \phi^{\ell}$ converges to 0. Also, $2\ell \phi^l \geq 0$ for $\ell\geq0$. For any $\epsilon$, we have $\ell'$ that satisfies $\epsilon \leq 2\ell\phi^\ell$ for $\ell' \leq \ell$. Then, fix such $\ell'$.
\begin{align*}
	&\int_{\ell'}^{\infty} \frac{4}{\pi} \text{arccotan}(\text{exp}\{\ell\}) dh < \infty \\
	& \sum_{\ell=\ell'}^{\infty}\frac{4}{\pi} \text{arccotan}(\text{exp}\{\epsilon/(2\phi^\ell)\} \leq \sum_{\ell = \ell'}^{\infty} \frac{4}{\pi} \text{arccotan}(\text{exp}\{\ell\})  < \infty 
\end{align*}
Therefore, 
\begin{align*}
	\sum_{\ell=0}^{\infty} \mathrm{P}(|z_{\ell}| \geq \epsilon) = \sum_{h=0}^{\infty}\frac{4}{\pi} \text{arccotan}(\text{exp}\{\epsilon/(2\phi^\ell)\} < \infty
\end{align*}
On the side note, we can also easily see that when $|\phi| >1$, the sum diverges to $\infty$. By Borel-Cantelli lemma, we may conclude that $\sum_{\ell=0}^{\infty} z_{\ell}$ diverges almost surely.

Define $y_{\ell} = z_{\ell}1_{\{|z_{\ell}| \leq \epsilon\}}$. For the second and the third condition, we need to show that 1) $\sum_{\ell=0}^{\infty} \mathbb{E}(y_{\ell})$ converges, and 2) $\sum_{\ell=0}^{\infty} \mathrm{Var}(y_{\ell})$, converges. For the second condition, it suffices to show:
$$\mathbb{E}(\sum_{\ell=0}^{\infty} y_{\ell})  = \sum_{\ell=0}^{\infty} \mathbb{E}(y_{\ell}) < \infty,$$
which is satisfied by showing $\sum_{\ell=0}^{\infty} \mathbb{E}(|y_{\ell}|) <\infty$ (Fubini). Using Cauchy-Schwartz Inequality:
\begin{align*}
	\sum_{\ell =0}^{\infty} \mathbb{E}(|y_{\ell}|) &= \sum_{\ell = 0}^{\infty} \mathbb{E}(|z_\ell 1_{\{|z_{\ell}| \leq \epsilon\}}|) \\
	&\leq \sum_{\ell=0}^{\infty} \sqrt{\mathbb{E}(z_{\ell}^2) \mathrm{P}(|z_{\ell}|\leq \epsilon)}\\
	&\leq \sum_{\ell=0}^{\infty} \sqrt{\mathbb{E}(z_{\ell}^2)} = \sum_{\ell=0}^{\infty}\sqrt{\mathrm{Var}(z_{\ell})} = \pi \sum_{\ell=0}^{\infty}\phi^{\ell} = \frac{\pi}{(1-\phi)} < \infty.
\end{align*}
The second condition is satisfied. For the third condition, similar logic is applied. First note that: 
\begin{align*}
	\mathrm{Var}\bigg(\sum_{\ell=0}^{\infty}y_{\ell}\bigg) &= \mathbb{E}\bigg(\bigg(\sum_{\ell=0}^{\infty}y_{\ell}\bigg)^2\bigg) + \bigg(\mathbb{E}\bigg(\sum_{\ell=0}^{\infty}y_{\ell}\bigg)\bigg)^2 \\
	&=\mathbb{E}\bigg(\bigg(\sum_{\ell=0}^{\infty}y_{\ell}\bigg)^2\bigg) + \bigg(\sum_{\ell=0}^{\infty}\mathbb{E}(y_{\ell})\bigg)^2.
\end{align*}
And, 
\begin{align*}
\lim_{n \to \infty}\mathbb{E}\bigg(\bigg(\sum_{\ell=0}^{n}y_{\ell}\bigg)^2\bigg)  &= \lim_{n \to \infty}\mathbb{E}\bigg(\sum_{\ell=0}^{n}y_{\ell}^2\bigg) & (independence) \\ 
&= \lim_{n \to \infty}\sum_{\ell=0}^{n}\mathbb{E}(y_{\ell}^2) = \sum_{\ell=0}^{\infty}\mathbb{E}(z_{\ell}^2 1_{\{|z_{\ell}| \leq \epsilon\}})\\
&\leq \sum_{\ell=0}^{\infty}\mathbb{E}(z_{\ell}^2) = \frac{\pi^2}{(1-\phi^2)} <\infty
\end{align*}
Therefore:
\begin{align*}
	\mathrm{Var}(\sum_{\ell=0}^{\infty}y_{\ell}) &= \sum_{\ell=0}^{\infty}\mathrm{Var}(y_{\ell}) < \infty
\end{align*}
All three conditions are satisfied.
\subsection{Theorem 2}\label{sec:proof_T3}
When $\phi = 0.5$, we have the following special case for the MGF, 
\begin{align*}
	\prod_{\ell=0}^{\infty}\cos(\frac{\pi t}{2^{\ell}}) &= \prod_{\ell=0}^{\infty}\frac{\sin(\frac{\pi t}{2^{\ell-1} })}{2\sin(\frac{\pi t}{2^{\ell}})} = \lim_{n \to \infty}\prod_{\ell=0}^{n}\frac{\sin(\frac{\pi t}{2^{\ell-1} })}{2\sin(\frac{\pi t}{2^{\ell}})}\\
	& = \lim_{n\to \infty} \frac{\sin(2\pi t) }{2^n \sin(\frac{\pi t }{2^n})} = \frac{\sin(2\pi t)}{2\pi t} 
\end{align*}
We have: 
\begin{align*}
&\prod_{\ell=0}^{\infty}\sec(\frac{\pi t}{2^{\ell}}) = \frac{1}{\prod_{\ell=0}^{\infty}\cos(\frac{\pi t}{2^{\ell}})} = \frac{2\pi t}{\sin(2\pi t)} = \Gamma(1-2t)\Gamma(1+2t) = \mathrm{B}(1-2t,1+2t)
\end{align*}
$(2\pi t)/\sin(2\pi t) = \Gamma(1-2t)\Gamma(1+2t)$ is by the reflection relation. $\text{Logistic}(\mu,s)$ has the the following moment generating function $\exp(-\mu t) \mathrm{B}(1-st,1+st)$. Since the moment generating function uniquely determines the random variable, and by Theorem 2, 
\begin{align*}
	z_t = \sum_{\ell = 0}^{\infty} z_{\ell,t} \stackrel{a.s}{\rightarrow} \text{Logistic}(0,2)
\end{align*}
\subsection{Theorem 3}\label{sec:proof_T4}
Define $\lambda_{t} := \exp(v_{t})$. We showed that $f(\lambda_{t}) = 1/(1+\lambda_{t})^2$. We have:
\begin{align*}
	f(\Delta h_t) = \int_{0}^{\infty} \frac{1}{\sqrt{2 \pi \lambda_{t}^2}}\text{exp}\bigg(-\frac{(\Delta h_t)^2}{2 \lambda_t^2}\bigg)\frac{1}{(1+\lambda_{t})^2} d \lambda_{t}
\end{align*}
$\forall x>0$,
	$$\frac{1}{2(1+x^{2})} \leq \frac{1}{(1+x)^2} \leq \frac{1}{(1+x^2)}.$$
Thus, 
\begin{align*}
	f(\Delta h_t) &\leq  \int_{0}^{\infty} \frac{1}{\sqrt{2 \pi \lambda_{t}^2}}\text{exp}\bigg(-\frac{(\Delta h_t)^2}{2 \lambda_t^2}\bigg)\frac{1}{(1+\lambda_{t}^2)} d \lambda_{t}\\ 
	&= \frac{1}{2\sqrt{2 \pi}} \int_{0}^{\infty} \frac{1}{1+u}\text{exp}\bigg(-\frac{u(\Delta h_t)^2}{2}\bigg)du & u = \frac{1}{\lambda_{t}^2}\\
	&= \frac{1}{2\sqrt{2 \pi}} \exp\{(\Delta h_t)^2/2\}E_1((\Delta h_t)^2/2),
\end{align*}
where $E_1()$ is the exponential integral function, which satisfies the following upper and lower bound $\forall t >0$:
$$ 
	\frac{\exp(-t)}{2} \text{log}\bigg(1 + \frac{2}{t}\bigg) < E_{1}(t) < \exp(-t) \text{log}\bigg(1 + \frac{1}{t}\bigg)
$$
Thus, $$ f(\Delta h_t) < \frac{1}{2\sqrt{2 \pi}} \text{log}\bigg(1 + \frac{2}{(\Delta h_t)^2}\bigg)$$
Similarly for the lower bound, 
$$ f(\Delta h_t) > \frac{1}{8\sqrt{2 \pi}} \text{log}\bigg(1 + \frac{4}{(\Delta h_t)^2}\bigg)
$$
The lower and upper bound for $\Delta h_t \neq 0$ is shown. The lower bound clearly approaches infinity as $\Delta h_{t}$ approaches 0, which completes the proof.
\subsection{Theorem 4}\label{sec:btfasv}
Let $\boldsymbol{\beta} := \{\beta_1, \ldots, \beta_T\}$ and $\boldsymbol{h} := \{h_1, \ldots, h_T\}$ denote the latent mean and log-volatility processes, respectively.  
For notational convenience, define $\boldsymbol{\lambda}_{\beta} := \{\sigma_{\beta,1}^2, \ldots, \sigma_{\beta,T}^2\}$ and $\boldsymbol{\lambda}_{h} := \{\sigma_{h,1}^2, \ldots, \sigma_{h,T}^2\}$. To prove identifiability of the hierarchical model, we need to establish that 
\begin{enumerate}
    \item For any $(\boldsymbol{h},\boldsymbol{\beta})$ and $(\boldsymbol{h}^*,\boldsymbol{\beta}^*)$ in the parameter space, if $f(\boldsymbol{y}|\boldsymbol{h},\boldsymbol{\beta}) = f(\boldsymbol{y}|\boldsymbol{h}^*,\boldsymbol{\beta}^*), \forall\boldsymbol{y} \in \mathcal{Y}$, then $(\boldsymbol{h},\boldsymbol{\beta}) = (\boldsymbol{h}^*,\boldsymbol{\beta}^*).$
    \item For any $(\boldsymbol{\lambda_{\beta}},\boldsymbol{\lambda_{h}})$ and $(\boldsymbol{\lambda_{\beta}}^*,\boldsymbol{\lambda_{h}}^*)$ in the hyperparameter space, if $f(\boldsymbol{y}|\boldsymbol{\lambda_{\beta}},\boldsymbol{\lambda_{h}}) = f(\boldsymbol{y}|\boldsymbol{\lambda_{\beta}}^*,\boldsymbol{\lambda_{h}}^*), \forall\boldsymbol{y} \in \mathcal{Y}$, then $(\boldsymbol{\lambda_{\beta}},\boldsymbol{\lambda_{h}}) = (\boldsymbol{\lambda_{\beta}}^*,\boldsymbol{\lambda_{h}}^*).$
\end{enumerate}
The first statement follows from the fact that, conditional on $(\boldsymbol{\beta},\boldsymbol{h})$, the likelihood is Multivariate Gaussian with mean $\boldsymbol{\beta}$ and with diagonal covariance matrix $\text{exp}\{\boldsymbol{h}\}$. Two parameters affect the density in functionally independent ways, thus the statement follows.

To prove the identifiability of the hyperparameter pair $(\boldsymbol{\lambda}_{\beta},\boldsymbol{\lambda}_{h})$, we proceed by contradiction.
Suppose there exist two distinct parameter vectors $(\boldsymbol{\lambda}_{\beta},\boldsymbol{\lambda}_h) \neq (\boldsymbol{\lambda}_{\beta}^*,\boldsymbol{\lambda}_h^*)$ such that $f(y|\boldsymbol{\lambda}_{\beta},\boldsymbol{\lambda}_h)=f(y|\boldsymbol{\lambda}_{\beta}^*,\boldsymbol{\lambda}_h^{*})$. 

By model specification $\boldsymbol{\beta}|\boldsymbol{\lambda}_{\beta} \sim N(\boldsymbol{0}, \mathbf{S}\Lambda_{\beta}\mathbf{S}^{T})$ and $\boldsymbol{h}|\boldsymbol{\lambda}_{h} \sim N(\boldsymbol{0}, \mathbf{S}\Lambda_h\mathbf{S}^{T})$ where $\Lambda_{\beta} = diag(\boldsymbol{\lambda}^2_{\beta})$, $\Lambda_{h} = \text{diag}(\boldsymbol{\lambda}^2_{h})$ and $\mathbf{S}$ is the un-differencing operator matrix corresponding to the differencing order k in $\Delta^k\beta$ and $\Delta^{k}h$.

Using the law of total variance, the variance of $y$ conditional on the hyperparameters can be decomposed as:
\begin{align*}
\mathrm{Var}(y \mid \boldsymbol{\lambda}_{\beta}, \boldsymbol{\lambda}_h)
&= E_{\boldsymbol{\beta}, \boldsymbol{h} \mid \boldsymbol{\lambda}_{\beta}, \boldsymbol{\lambda}_h} 
\left[ \mathrm{Var}(y \mid \boldsymbol{\beta}, \boldsymbol{h}) \right] + \mathrm{Var}_{\boldsymbol{\beta}, \boldsymbol{h} \mid \boldsymbol{\lambda}_{\beta}, \boldsymbol{\lambda}_h}
\left[E(y \mid \boldsymbol{\beta}, \boldsymbol{h}) \right] \\
&= E_{ \boldsymbol{h} \mid \boldsymbol{\lambda}_h} [\text{exp}\{\boldsymbol{h}\}] + \mathrm{Var}_{\boldsymbol{\beta}|\lambda_{\beta}}[\boldsymbol{\beta}]\\
&= \text{exp}\bigg\{\frac{1}{2}\text{diag}(\mathbf{S}\Lambda_{h}\mathbf{S}^{T})\bigg\} + \mathbf{S}\Lambda_{\beta}^2\mathbf{S}^{T}.
\end{align*}
The first term is diagonal, while the second term is dense due to the cumulative sum operation of $\mathbf{S}$. 

The assumption of equal marginal distributions implies equality of variances:
\begin{align*}
  &\text{exp}\bigg\{\frac{1}{2}\text{diag}(\mathbf{S}\Lambda_{h}\mathbf{S}^{T})\bigg\} + \mathbf{S}\Lambda_{\beta}\mathbf{S}^{T} =   \text{exp}\bigg\{\frac{1}{2}\text{diag}(\mathbf{S}\Lambda_{h}^*\mathbf{S}^{T})\bigg\} + \mathbf{S}\Lambda_{\beta}^*\mathbf{S}^{T},\\
  &\text{exp}\bigg\{\frac{1}{2}\text{diag}(\mathbf{S}\Lambda_{h}\mathbf{S}^{T})\bigg\} -
  \text{exp}\bigg\{\frac{1}{2}\text{diag}(\mathbf{S}\Lambda_{h}^*\mathbf{S}^{T})\bigg\}
  = \mathbf{S}\Lambda_{\beta}^*\mathbf{S}^{T} - 
  \mathbf{S}\Lambda_{\beta}\mathbf{S}^{T}.
\end{align*}
The left hand side is always diagonal while the right hand side is dense unless $\boldsymbol{\lambda}_{\beta}=\boldsymbol{\lambda}_{\beta}^*$. Hence, the only solution consistent with both sides is when both are zero matrix, implying $\boldsymbol{\lambda}_{\beta}=\boldsymbol{\lambda}_{\beta}^*$ and $\boldsymbol{\lambda}_{h}=\boldsymbol{\lambda}_{h}^*$. This is a contradiction to the assumption of distinct parameters: $(\boldsymbol{\lambda}_{\beta},\boldsymbol{\lambda}_h) \neq (\boldsymbol{\lambda}_{\beta}^*,\boldsymbol{\lambda}_h^*).$ Therefore, the parameters $(\boldsymbol{\lambda}_{\beta},\boldsymbol{\lambda}_{h})$ are identifiable.
\section{Full Conditional for Gibbs Sampling}\label{App:A}
\subsection{\texorpdfstring{$j$}{Lg}}\label{sec:j}
$\boldsymbol{j} = (j_1,\ldots,j_T)$ was introduced to expand the likelihood on $\boldsymbol{y^*} := (\log(y_1^2),\ldots,\log(y_{T}^2))$. In this section, we show that $\forall k \in \{1,\ldots,10\}$:
\begin{align*}
\mathrm{p}(j_t=k| h_t,v_t,s_t,\xi_t,\mu,\xi_{\mu},\phi,y_t^*)  &= \mathrm{p}(j_t = k| h_t,y_t^*) \\
    &= \frac{\mathcal{N}(y_t^*| h_t + m_{k},w_{k}^2) p_k}{\sum_{i = 1}^{10} p_i\mathcal{N}(y_t^*| h_t + m_{i},w_{i}^2) },&\forall t \in \{1,\ldots,T\}
\end{align*}
$\boldsymbol{j}$ is only associated with $y^*$, which is only associated with $\boldsymbol{h}$ and $\boldsymbol{j}$:
\begin{align*}
\mathrm{p}(j_t = k | h_t,v_t,s_t,\xi_t,\mu,\xi_{\mu},\phi,y_t^*) &= \mathrm{p}(j_t=k| h_t,y_t^*) = \frac{f(y_t^*| h_t,j_t)\mathrm{p}(j_t=k)}{\int f(y_t^*| h,j_t)f(j_t) dj_t}
\end{align*}
By \citet{omorietal}, 
\begin{align*}
f(y_t^*|j_t, h_t) =\mathcal{N}(y_t^*| h_t + m_{j_t}, w_{j_t}^2)  
\end{align*}
Naturally, $\forall k \in \{1,\ldots,10\}$
\begin{align*}
\mathrm{p}(j_t= k | h_t,y_t^*) &=\frac{f(y^*_t| h_t,j_t = k)\mathrm{p}(j_t = k)}{f(y^*_t | h_t)}\\
&= \frac{f(y^*_t| h_t,j_t = k)\mathrm{p}(j_t=k)}{\sum_{i = 1}^{10}\mathrm{p}(j_t = i) f(y^*_t| h_t,j_t = i)}\\
&= \frac{\mathcal{N}(y_t^*| h_t + \mu_{k},\sigma_{k}^2) p_k}{\sum_{i = 1}^{10} p_i\mathcal{N}(y_t^*| h_t + \mu_{i},\sigma_{i}^2) }, &\forall t \in \{1,\ldots,T\},
\end{align*}
which is what we wanted to show. Exact distribution on $j_t\stackrel{i.i.d}{\sim}  \text{Categorical}(\pi^{\text{Omori}})$ as well as corresponding mean and the variance parameter of each component is described in \cite{omorietal}.
\subsection{\texorpdfstring{$ h$}{Lg}}\label{sec:h}
The likelihood and the conditional priors on $\boldsymbol{h}$ are:
\begin{align*}
    f(y^*|\boldsymbol{j}, \boldsymbol{h}) &= \prod_{t=1}^{T} \mathcal{N}(y_t^*| h_t + m_{j_t}, w_{j_t}^2) = \mathcal{N}(y^*| h + m_j, w_{j}^2I )\\
    f(\boldsymbol{h}|\textbf{v}) & = \mathcal{N}( h_1|0,e^{v_1})\prod_{t=1}^{T}\mathcal{N}( h_t| h_{t-1},e^{v_t})
\end{align*}
Note that $\boldsymbol{h}$, a conditionally Gaussian, is a linear combination of $\Delta \boldsymbol{h}$, with the cumulative sum operator, the precision of the conditional prior of $h$ is a tridiagonal matrix $Q_v$ 
\[
Q_v := 
\begin{bmatrix}
     \bigg(\frac{1}{e^{v_2}}+\frac{1}{e^{v_1}}\bigg) & -\frac{1}{e^{v_2}} &0 & \ldots & \ldots & 0\\
     -\frac{1}{e^{v_2}}  & \bigg(\frac{1}{e^{v_3}}+\frac{1}{e^{v_2}}\bigg) & -\frac{1}{e^{v_3}} &  \ddots &  \ddots & \vdots\\
     0 & -\frac{1}{e^{v_3}}  & \bigg(\frac{1}{e^{v_4}}+\frac{1}{e^{v_3}}\bigg) & -\frac{1}{e^{v_4}} &  \ddots &  \ddots \\
     \vdots& \ddots  & \ddots & \ddots & \ddots & 0\\
     \vdots       &\ddots   & \ddots  & -\frac{1}{e^{v_{T-1}}} & \bigg(\frac{1}{e^{v_{T-1}}}+\frac{1}{e^{v_T}}\bigg) & -\frac{1}{e^{v_{T}}}\\
     0 & \ldots & \ldots & 0 & -\frac{1}{e^{v_T}} & \frac{1}{e^{v_T}} \\
\end{bmatrix}
\]
Since both the likelihood and the prior are Gaussian, the conditional posterior is also Gaussian:
\begin{align*}
f( \boldsymbol{h}|\boldsymbol{j},\boldsymbol{v},\boldsymbol{y^*}) = \mathcal{N}\bigg(\boldsymbol{h}\bigg|\bigg(Q_v + I\frac{1}{w^2_{\boldsymbol{j}}}\bigg)^{-1}\frac{y-m_{\boldsymbol{j}}}{w^2_{\boldsymbol{j}}},\bigg(Q_v + I\frac{1}{w^2_{\boldsymbol{j}}}\bigg)^{-1} \bigg).
\end{align*}
\subsection{\texorpdfstring{$v$}{Lg}}\label{sec:v}
We use the all-without-loop (AWOL) sampler by \cite{Kastner_2014} with parameter expansion on the error term with scale mixture normal distribution for sampling $\textbf{v},\mu$ and $\phi.$ By the conditional independence, the likelihood with respect to $\boldsymbol{v}$, reduces to $f(\boldsymbol{\omega^*}|v)$ where $\omega_1^* = \text{log}( h_1^2)$ and $\omega_t^* = \text{log}(( h_t -  h_{t-1})^2), \forall t\geq 2.$ With the parameter expansion on the likelihood with 10-component Gaussian Mixture proposed by \cite{omorietal},
\begin{align*}
    f(\boldsymbol{\omega}^*|\textbf{v},\boldsymbol{s}) &= \mathcal{N}(\omega^*|v+m_s,w_{s}^2I).
\end{align*}
The conditional prior distribution of $v$ is also Gaussian due to the Z-distribution being a scale mixture Gaussian. $\boldsymbol{v}$ is also a linear combination of $\boldsymbol{v}^* = (v^*_{1},\ldots,v^*_{T})'$, where $v_1^* = v_1$ and $v^*_{t+1} = v_{t+1} - \phi v_{t}, t\geq 2$, conditionally independent Gaussian random variables:
\begin{spreadlines}{2ex}
\begin{align*}
    f(v^*_1|\xi_1,\mu) &= f(v_1|\xi_0,\mu) = \mathcal{N}(v^*_{t}|\mu,1/\xi_1)\\
    f(v^*_{t}|\xi_{t-1},\mu,\phi) &= f(v_t - \phi v_{t-1}|\xi_{t-1},\mu,\phi) = \mathcal{N}(v^*_{t}|\mu (1-\phi),1/\xi_{t-1}) & t \geq 2,
\end{align*}
\end{spreadlines}
Similar to section~\ref{sec:h}, we have:
\[
    f(\boldsymbol{v}|\boldsymbol{\xi},\mu,\phi) = \mathcal{N}(\boldsymbol{v}|0,Q_{\xi,\phi}^{-1}),
\]
with
\[Q_{\xi,\phi} = 
\begin{bmatrix}
     \xi_1+\phi^2\xi_2 & -\phi\xi_2 &0 & \ldots & \ldots & 0\\
     -\phi\xi_2  &\xi_2+\phi^2\xi_3 & -\phi\xi_3 &  \ddots &  \ddots & \vdots\\
     0 & -\phi \xi_3  & \xi_3+\phi^2\xi_4 &  -\phi\xi_4 &  \ddots \\
     \vdots& \ddots  & \ddots & \ddots & \ddots & 0\\
     \vdots       &\ddots   & \ddots  & -\phi\xi_{T-1} & \xi_{T-1}+\phi^2\xi_{T}  & -\phi\xi_{T}\\
     0 & \ldots & \ldots & 0 & -\phi\xi_{T} & \xi_{T} \\
\end{bmatrix}
\]
Thus, 
\begin{align*}
    f(\textbf{v}|y^*,\ldots) &= 
    \mathcal{N}\bigg(\textbf{v}\bigg|\bigg(Q_{\xi,\phi} + I\frac{1}{w^2_{\boldsymbol{s}}}\bigg)^{-1}\bigg(\frac{\boldsymbol{\omega^*}- m_{\boldsymbol{s}}}{w^2_{\boldsymbol{s}}} + Q_{\xi,\phi} \boldsymbol{1} \mu\bigg), \bigg(Q_{\xi,\phi} + I\frac{1}{w^2_{\boldsymbol{s}}}\bigg)^{-1} \bigg)
\end{align*}

\subsection{\texorpdfstring{$s$}{Lg}}\label{sec:s}
In section~\ref{sec:h}, $\boldsymbol{s} = (s_1,\ldots,s_{T})'$ was introduced to expand $\boldsymbol{\omega}^*|\textbf{v}$. Based on the same argument used in section~\ref{sec:j}, $\forall k \in \{1,\ldots,10\}$
\begin{align*}  
    \mathrm{p}(s_t=k|\omega_t^*,v_t) &= \frac{\mathcal{N}(\omega_t^*|v_t + m_{k},w_{k}^2) p_k}{\sum_{i = 1}^{10} p_i\mathcal{N}(\omega^*_t|v_t + m_{i},w_{i}^2) }, &\forall t \in \{1,\ldots,T\}.
\end{align*}
\subsection{\texorpdfstring{$\xi$}{Lg}}\label{sec:xi}
Based on \cite{polyagamma}, we have
\begin{align*}
    v_1 & = \mu + \eta_{0} \\ 
    v_{t} &= \mu + \phi(v_{t-1}-\mu) + \eta_{t-1}, &\forall t \geq 2.
\end{align*}
Naturally, $\eta_1 = v_1 - \mu$ and $\eta_t = v_{t+1}-\phi v_{t} - \mu(1-\phi), \forall t \geq 2$. 
\begin{align*}
    f(\xi_0|\textbf{v},\mu,\phi) &= \mathcal{PG}(\xi_0|1,v_1 - \mu)\\
    f(\xi_{t-1}|\textbf{v},\mu,\phi) &= \mathcal{PG}(\xi_{t-1}|1,v_t - \phi v_{t-1} - \mu(1-\phi)) &\forall t \geq 2,
\end{align*}
where $\mathcal{PG}$ represents the density function for P{\'o}lya-Gamma random variable.
\subsection{\texorpdfstring{$\mu$}{Lg}}\label{sec:mu}
Define $\hat{v}_{\phi}^* = \frac{\sum_{t=2}^{T}\sqrt{\xi_t}(v_t - \phi v_{t-1})}{(1-\phi)\sum_{t=2}^{T}\sqrt{\xi_{t}}}$, so that 
\begin{align*}
    &v_t - \phi v_{t-1}|\boldsymbol{\xi},\mu,\phi \sim N((1-\phi)\mu, \frac{1}{\xi_{t}}) \\
    &\hat{v}_{\phi}^*|\boldsymbol{\xi},\mu,\phi \sim N(\mu,\sigma^2_{\xi,\phi})\\
    &\sigma^2_{\xi,\phi}  = \frac{T-1}{((1-\phi)\sum_{t=2}^{T}\sqrt{\xi_{t}})^2}.
\end{align*}
The conditional prior distribution on $\mu$ also reduces to $f(\mu|\xi_{\mu}) = \mathcal{N}(\mu|0,\frac{1}{\xi_{\mu}})$ Thus, the conditional posterior distribution for $\mu$ is
\begin{align*}
f(\mu|\hat{v}^*_{\phi},\boldsymbol{\xi},\xi_{\mu},\phi) &= \mathcal{N}\bigg(\mu\bigg|\bigg(\frac{1}{\sigma^2_{\xi,\phi}} + \xi_{\mu} \bigg)^{-1}\frac{\hat{v}^*_{\phi}}{\sigma^2_{\xi,\phi}}, \bigg(\frac{1}{\sigma^2_{\xi,\phi}} + \xi_{\mu} \bigg)^{-1}\bigg).
\end{align*}
\subsection{\texorpdfstring{$\xi_{\mu}$}{Lg}}\label{sec:ximu}
Based on the scale mixture representation of the Z-distribution \cite{polyagamma}:
\begin{align*}
     f(\xi_{\mu}|\mu)= \mathcal{PG}(\xi_{\mu}|1,\mu).
\end{align*}
\subsection{\texorpdfstring{$\phi$}{Lg}}\label{sec:phi}
Prior distribution is imposed on $\phi^* := (\phi + 1)/2$, where $\phi^* \sim \text{Beta}(1/2,1/2)$. $\phi = 2\phi^* - 1$. We may rewrite the equation as a regression form:
\begin{align*}
    &v_{t} = \mu + \phi(v_{t-1} - \mu) + \eta_t ,\\
    &v_{t} = \mu + (2\phi^* - 1)(v_{t-1} - \mu) + \eta_t ,\\
    &(v_t + v_{t-1} - 2\mu) = \phi^*(2v_{t-1} - 2\mu) + \eta_t.
\end{align*}
Because $\eta_t$ is conditionally normal, the formulation can be viewed as a simple linear regression where the dependent variable is $v_t + v_{t-1} - 2\mu$, the regressor is $(2v_{t-1} - 2\mu)$, and the regression coefficient $\phi^*$. Conditional on all other parameters, the likelihood for $\phi^*$ is therefore univariate Gaussian. Combining this likelihood with the non-conjugate, but exponential-family $\text{Beta}(1/2,1/2)$ prior yields a smooth posterior over $(0,1)$, which is sampled using slice sampling algorithm by \citet{slice}.
\section{Simulation Set-Up}
\begin{table}[!htbp]
\centering 
\resizebox{0.8\textwidth}{!}{
\begin{tabular}[t]{|l|l|}
  \hline
  \textbf{DGP1: SV with 1 Regime}                   &
  \textbf{DGP2: SV with 2 Regimes}                  \\ \hline                      
  \parbox{80pt}{
    \begin{align*}
      &y_t \sim N(0,\exp(h_t))\\
      &h_{t}= -7.89 + 0.94(h_{t-1} - 7.89)+ 0.327u_{t}\\
      &u_t\stackrel{iid}{\sim}N(0,1)
    \end{align*}
  }                                                  &
  \parbox{80pt}{
    \begin{align*}
    &y_t \sim N(0,\exp(h_t))\\
    & h_{t} = m_{s_{t}}  + 0.85(h_{t-1} - m_{s_{t-1}}) + 0.0461 u_{t} \\ 
    &u_t\stackrel{iid}{\sim}N(0,1)\\
	& m_{s_t} =\begin{cases}
			\, 3.35\, , &\text{if $s_t = 0$.} \\
			\, 10.47 \, , &\text{if $s_t = 1$.}    
			\end{cases}
    \end{align*}
  }	                                                 \\ \hline
  \textbf{DGP3: SV with 3 Regimes}                  &
  \textbf{DGP4: GARCH with 1 Regime}                 \\ \hline
  \parbox{80pt}{
    \begin{align*}
    &y_t \sim N(0,\exp(h_t))\\
    & h_{t} = m_{s_{t}}  + 0.65(h_{t-1} - m_{s_{t-1}}) + 0.065 u_{t} \\ 
    &u_t\stackrel{iid}{\sim}N(0,1)\\
	& m_{s_t} =\begin{cases}
			\, 1.46 \, , &\text{if $s_t = 0$.} \\
			\, 3.91 \, , &\text{if $s_t = 1$.}    \\
			\, 6.22 \, , &\text{if $s_t = 2$.}    
			\end{cases}
    \end{align*}
  }		&
  \parbox{80pt}{
    \begin{align*}
    &y_t \sim N(0,\sigma^2_{y,t})\\
     &\sigma_{y,t}^{2} = 0.001 + 0.263 y_{t-1}^2 + 0.705 \sigma_{y,t-1}^2
    \end{align*}
  }                                                   \\ \hline
  \textbf{DGP5: GARCH with 2 Regimes}                  &
  \textbf{DGP6: GARCH with 3 Regimes}                 \\ \hline 
  \parbox{80pt}{
    \begin{align*}
        & y_t \sim N(0,\sigma^2_{y,t})\\
	& \sigma_{y,t}^{2} = m_{s_{t}} + \alpha_{s_t} y_{t-1}^2  + \beta_{s_{t}} \sigma_{y,t-1}^2\\
	& m_{s_t} =\begin{cases}
			\, 1.671 \, , &\text{if $s_t = 0$.} \\
			\, 108.67\, , &\text{if $s_t = 1$.}    
			\end{cases} \\
	& \alpha_{s_{t}} =\begin{cases}
			\, 0.0003 \, , &\text{if $s_t = 0$.} \\
			\, 0.0181\, , &\text{if $s_t = 1$.}    
			\end{cases}\\
	& \beta_{s_{t}} =\begin{cases}
			\, 0.99 \, , &\text{if $s_t = 0$.} \\
			\, 0.9042\, , &\text{if $s_t = 1$.}    
			\end{cases}
	\end{align*}
  }  &                                                
  \parbox{80pt}{
    \begin{align*}
        & y_t \sim N(0,\sigma^2_{y,t})\\
	& \sigma_{y,t}^{2} = m_{s_{t}} + \alpha_{s_t} y_{t-1}^2  + \beta_{s_{t}}\sigma_{y,t-1}^2\\
	& m_{s_t} =\begin{cases}
			\, 0.002 \, , &\text{if $s_t = 0$.} \\
			\, 0.051\, , &\text{if $s_t = 1$.} \\
			\, 13.35\, , &\text{if $s_t = 2$.}    
			\end{cases} \\
	& \alpha_{s_{t}} =\begin{cases}
			\, 0.073 \, , &\text{if $s_t = 0$.} \\
			\, 0.0035\, , &\text{if $s_t = 1$.} \\   
			\, 0.102\, , &\text{if $s_t = 2$.} 
			\end{cases}\\
	& \beta_{s_{t}} =\begin{cases}
			\, 0.927 \, , &\text{if $s_t = 0$.} \\
			\, 0.906\, , &\text{if $s_t = 1$.} \\   
			\, 0.895\, , &\text{if $s_t = 2$.}    
			\end{cases}
	\end{align*}
  }                                                  \\ \hline
  \textbf{DGP7: Sinusoidal}                  &
  \textbf{DGP8: Piecewise Constant}                 \\ \hline 
  \parbox{80pt}{
    \begin{align*}
    &y_t \sim \mathcal{N}(0, \exp(h_t))\\
    &h = A\sin(10(2\pi t)/T) + B\cos(10(2\pi t)/T)  \\ 
    &\quad+ C\sin(3(2\pi t)/T) + D\cos(3(2\pi t)/T)\\
    &A,B,C,D\stackrel{ind}{\sim}U(0,5)
	\end{align*}
  }  &                                                
  \parbox{80pt}{
\begin{align*}
    &y_t \sim \mathcal{N}(0, \exp(h_t))\\
    &j := \left\lfloor \frac{t}{25} \right\rfloor + 1\\
    &h_t = (-1)^j \cdot |z_j|\\
    &z_j \sim \begin{cases}
        N(5,0.5^2) & \text{if $j$ is even}\\
        N(0,0.5^2) & \text{if $j$ is odd}
    \end{cases}
\end{align*}
  }                                                  \\ \hline
\end{tabular}}
\caption{Summary of the eight data‐generating processes (DGPs). DGPs 1–3 simulate stochastic‐volatility models with one, two, or three regimes, and DGPs 4–6 simulate GARCH(1,1) models with one, two, or three regimes. In DGP 7, $h_{t}$ is a random sinusoid, and in DGP 8, $h_{t}$ is piecewise constant with random intercepts. For DGPs 2 and 4, the transition probability of remaining in the same state is 0.98, and transitioning to a different state is 0.02. For DGPs 3 and 6, the probability of staying in the same state is 0.98, while switching to any other state occurs with probability 0.01. Each DGP comprises 1,000 simulated sample paths, each consisting of 1,000 observations.
}
\label{tab:DGP}
\end{table}

\end{document}